\documentclass[twocolumn,trackchanges]{aastex701}

\usepackage{hyperref}
\begin{document}

\title{A Newly Identified Transient Ultraluminous X-ray Source in M31}

\author[orcid=0009-0008-7581-3096,sname='Ferdinand']{Ferdinand}
\affiliation{Cahill Center for Astrophysics, California Institute of Technology, MC 249-17, 1216 E California Boulevard, Pasadena, CA, 91125, USA}
\affiliation{Department of Astronomy, University of Illinois at Urbana-Champaign, Urbana, IL 61801, USA}
\email[show]{ff10@illinois.edu}  

\author[orcid=0000-0002-8147-2602, gname='Murray', sname='Brightman']{Murray Brightman} 
\affiliation{Cahill Center for Astrophysics, California Institute of Technology, MC 249-17, 1216 E California Boulevard, Pasadena, CA, 91125, USA}
\email{mbright@caltech.edu}

\author[orcid=0000-0002-6412-0103, gname='Jean-Marie', sname='Hameury']{Jean-Marie Hameury} 
\affiliation{Université de Strasbourg, CNRS, Observatoire astronomique de Strasbourg, UMR7550, 67000 Strasbourg, France}
\email{jean-marie.hameury@astro.unistra.fr}

\author[orcid=0000-0002-0273-218X]{Elias Kammoun}
\affiliation{Cahill Center for Astrophysics, California Institute of Technology, MC 249-17, 1216 E California Boulevard, Pasadena, CA, 91125, USA}
\email{ekammoun@caltech.edu}

\author[orcid=0000-0001-5819-3552]{Dominic J. Walton}
\affiliation{Centre for Astrophysics Research, University of Hertfordshire, College Lane, Hatfield AL10 9AB, UK}
\email{d.walton4@herts.ac.uk}

\author[orcid=0000-0002-8183-2970, sname=Middleton, gname=Matthew]{Matthew J. Middleton}
\affiliation{Department of Physics and Astronomy, University of Southampton, Highfield, Southampton SO17 1BJ, UK}
\email{M.J.Middleton@soton.ac.uk}

\author[orcid=0000-0001-5857-5622, sname=Earnshaw,gname=Hannah]{Hannah Earnshaw}
\affiliation{Cahill Center for Astrophysics, California Institute of Technology, MC 249-17, 1216 E California Boulevard, Pasadena, CA, 91125, USA}
\email{hpearn@caltech.edu}

\author[orcid=0000-0003-0388-0560]{Felix F\"urst}
\affiliation{European Space Agency (ESA), European Space Astronomy Centre (ESAC), Camino Bajo del Castillo s/n, 28692 Villanueva de la Cañada, Madrid, Spain}
\email{Felix.Fuerst@esa.int}

\author[orcid=0000-0002-3934-2644, sname=Jacobson-Galán,gname=Wynn]{Wynn V. Jacobson-Galán}
\altaffiliation{NASA Hubble Fellow}
\affiliation{Cahill Center for Astrophysics, California Institute of Technology, MC 249-17, 1216 E California Boulevard, Pasadena, CA, 91125, USA}
\email{wynnjg@caltech.edu}

\author[orcid=0000-0001-5754-4007]{Jacob E. Jencson}
\affiliation{Caltech/IPAC, Mail Code 100-22, 1200 E. California Blvd., Pasadena, CA 91125, USA}
\email{jjencson@ipac.caltech.edu}

\author[0000-0003-3252-352X]{Margaret Lazzarini}
\affiliation{Department of Physics \& Astronomy, California State University Los Angeles, 5151 State University Drive, Los Angeles, CA 90032, USA}
\email{mlazzar2@calstatela.edu}

\author[0000-0003-1621-9392]{Mark T. Reynolds}
\affiliation{Department of Astronomy, Ohio State University, 140 West 18th Ave., Columbus, OH 43210, USA}
\affiliation{Center for Cosmology and AstroParticle Physics, 191 W Woodruff Ave, Columbus, OH 43210, USA}
\email{reynolds.1362@osu.edu}

\begin{abstract}

Ultraluminous X-ray sources (ULXs) are off-nuclear X-ray sources with luminosities exceeding the Eddington limit of typical stellar-mass compact objects. While most known ULXs are relatively persistent sources, transient ULXs have emerged as a distinct population that offer unique insight into extreme, super-Eddington accretion. We report the discovery and X-ray characterization of a newly identified transient ULX in M31. First detected serendipitously on 2021 June 9, the source reached a peak unabsorbed $0.3 - 10$ keV luminosity of $\sim10^{39}\ \rm erg\ s^{-1}$. Its spectral evolution suggests increasing advection at later epochs, opposite to the expected transition from a super-Eddington slim disk toward a standard thin disk during the decay of the outburst. The observed X-ray light curve can be reproduced using the disk-instability model of Hameury \& Lasota, with characteristic timescales of $\sim41-61$ days. The inferred mass-accretion rate is super-Eddington for a $1.4\,M_{\odot}$ neutron star but approximately Eddington for a $10\,M_{\odot}$ stellar-mass black hole. Nearby optical sources from the \textit{Hubble Source Catalog} suggest a possible giant donor with an age of $\log_{10}({\rm age/yr})\sim8.5$ and a mass of $\sim3.2\,M_{\odot}$, when compared with stellar isochrones with an assumed $A_V=1.58$. We estimate a transient ULX occurrence rate in M31 of $0.18\pm0.10\ {\rm yr^{-1}}$. This rate is lower than those reported for other galaxies in previous studies. M31 ULX-3 expands the small population of known transient ULXs in M31 and highlights the importance of continued, regular X-ray monitoring of nearby galaxies.

\end{abstract}


\section{Introduction} \label{sec:introduction}

Ultraluminous X-ray sources (ULXs) are off-nuclear X-ray sources whose apparent luminosities exceed $10^{39}\ \rm erg\ s^{-1}$. At these luminosities, ULXs radiate near or above the Eddington limit expected for a $\sim10\,M_{\odot}$ black hole \citep[see][]{Kaaret2017,Fabrika2021,King2023, Pinto2023}. First discovered in 1980s by the Einstein Observatory \citep{Giacconi1797}, early interpretations about ULXs were intermediate-mass black holes (IMBHs) with masses of approximately $10^{2}-10^{5}\,M_{\odot}$ \citep[e.g.,][]{Colbert1999, Strohmayer2009}. However, the spectral and timing properties of many ULXs are now more commonly interpreted in terms of super-Eddington accretion onto stellar-mass compact objects \citep[e.g.,][]{Gladstone2009,Takeuchi2013,Bachetti2013,Middleton2015, Kosec2018a, Kosec2018b}. The discovery of coherent X-ray pulsations in several systems provided direct evidence that some ULXs contain neutron stars \citep[e.g.,][]{Bachetti2014,Furst2016,Israel2017a,Israel2017b, Carpano2018, Sathyaprakash2019, Castillo2020, Pintore2025, Ducci2025}. Because neutron stars have masses of only $\sim1-2\,M_{\odot}$, their observed luminosities may exceed the corresponding Eddington limit by more than two orders of magnitude $(\rm up \ to \ L_X\sim 10^{41} \ erg \ s^{-1})$. ULXs consequently become some of the best laboratories for studying super-Eddington accretion.

Although most well-studied ULXs remain luminous over long timescales \citep[with some having estimated dynamical ages of $10^5$ years based on the dynamics of the nebulae in which they sit;][]{Pakull2002, Pakull2010, Moon2011, Weng2014, Berghea2020, Soria2021, Gurpide2022}, many ULXs have been found to exhibit transient behavior. \citet{Brightman2023} argued that transient systems may even dominate the underlying population of sources capable of reaching ULX luminosities, while persistent ULXs dominate the bright end of the galaxy X-ray luminosity functions. These transient ULXs are generally undetected or substantially fainter during quiescence and brighten above $10^{39}\ \rm erg\ s^{-1}$ during outbursts lasting from weeks to months, or in some cases longer \citep[e.g.,][]{Middleton2012_ULX1, Barnard2013_M31ULX2, Earnshaw2018, Dage2021, Walton2021, Brightman2023}. Transient ULX behavior may arise from several types of accreting systems, with most proposed scenarios involving neutron stars or black holes in X-ray binaries (XRBs).

One possible origin involves Be/X-ray binaries (Be XRBs), often containing strongly magnetized neutron stars, in which enhanced accretion from the decretion disk of a Be-star companion produces giant outbursts that temporarily reach the ULX regime \citep{Reig2011}. There are several transient ULX examples originating from Be XRBs, including the Galactic transient ULX, Swift~J0243.6+6124 \citep{Cenko2017, Wilson-Hodge2018}, and a transient ULX from the SMC, SMC~X-3 \citep{Tsygankov2017}. However, Be XRBs are not the only one to account for the full transient ULX population \citep{Brightman2023}. Other transient ULXs may instead be low-mass X-ray binaries (LMXBs) containing stellar-mass black holes or neutron stars. In these systems, thermal-viscous instabilities in the accretion disk can trigger outbursts during which the accretion rate becomes sufficiently high for the source to enter the ULX regime \citep{Lasota2001,Hameury2020}.

Alternatively, transient ULX may arise from transitions into the propeller regime. When a magnetized neutron star accretor is approaching spin equilibrium, the co-rotation radius can approach the magnetospheric radius, partially or completely inhibiting accretion onto the neutron star surface \citep[see][]{Illarionov1975,Tsygankov2016a,Tsygankov2016b, Earnshaw2018,Middleton2023}. This transition from accretion to the propeller state can produce a sharp decrease in X-ray luminosity \citep{Dall2015,Furst2016,Tsygankov2016a,Earnshaw2018}.

Transient ULX can also be produced geometrically through precession of a beamed supercritical accretion flow. As the geometrically beamed wind cone precesses, the beam can move into and out of our line of sight, causing substantial changes in the observed luminosity and potentially producing transient ULX behavior \citep[see][and references therein]{Khan2022}. Such precession may be induced by several external torques, including tidal eﬀects, radiation pressure driven instabilities, Lense--Thirring or solid body precession, magnetic warping and free-free precession \citep{Pringle1996,Maloney1997,Maloney1998,Fragile2007,Lei2013,Middleton2018, Middleton2019}.

% Andromeda Galaxy
A particularly well-monitored nearby galaxy is M31 (the Andromeda galaxy). Over the past two decades of X-ray observations, two ULXs have been discovered and characterized in M31: M31 ULX-1 \citep[CXOM31 J004253.1+411422;][]{Middleton2012_ULX1, Straub2013} and M31 ULX-2 \citep[XMMU J004243.6+412519;][]{Barnard2013_M31ULX2, Esposito2013, Middleton2013}. Interestingly, both well-established ULXs in M31 are transient, and no persistent ULX has yet been securely identified in the galaxy. M31 ULX-1 was discovered in 2009 with a peak X-ray luminosity of $\sim 5 \times 10^{39}\ \rm erg\ s^{-1}$ and declined on an e-folding timescale of $\sim40$ days. Its thermal, disk-like spectrum, resemblance to Galactic black-hole transients, and lack of a luminous O/B-type companion support its interpretation as a stellar-mass black hole in an LMXB \citep{Middleton2012_ULX1,Straub2013}. Meanwhile, M31 ULX-2 was discovered in 2012 with a peak X-ray luminosity of $\sim10^{39}\ \rm erg\ s^{-1}$. It remained at a relatively stable luminosity for at least $\sim40$ days before fading below $10^{38}\ \rm erg\ s^{-1}$ over the following $\sim200$ days. Similar to M31 ULX-1, the spectral evolution of M31 ULX-2 resembled that of Galactic black-hole transients, while the fading of its optical counterpart provided evidence for a low-mass donor \citep{Barnard2013_M31ULX2, Esposito2013, Middleton2013}. Together, these two systems suggest that transient ULXs in M31 may be stellar-mass black-hole LMXBs temporarily reaching super-Eddington luminosities during bright accretion outbursts.

\citet{Brightman2023} discovered five new transient ULXs in nearby galaxies, mostly in real time, finding that they are remarkably common. Motivated by this, and to expand the sample further, we searched through archival \textit{Swift X-ray Telescope} (\textit{Swift}/XRT) observations of nearby galaxies to find further such sources. %As in \citet{Brightman2023}, we searched for sources in these observations using the {\tt detect} function of the {\tt HEASoft} \citep{HEASoft2014} tool {\tt ximage} and a signal to noise threshold of 3. For each source detected we examined their light curves for transient behaviour. 
One such source was LSXPS J004321.0+411833, a previously uncharacterized transient associated with M31. It was first detected in 2021 June 9 with \textit{Swift}/XRT at a count rate of $0.44 \pm 0.04 \ \rm ct \ s^{-1}$ (ObsID: 00013977006), corresponding to an X-ray luminosity of $\sim10^{39}\ {\rm erg\ s^{-1}}$. Given its high luminosity and its projected offset of more than $2'$ from the center of M31, it can be classified as a ULX. Since this is the third ULX in M31, we will refer to this source as M31 ULX-3.

We present and analyze the X-ray observations of M31 ULX-3 from the \textit{Chandra X-ray Observatory} (CXO) and the \textit{Swift}/XRT. M31 ULX-3 is located at $(\alpha = \rm 0^h43^m20.93^s, \delta=41^{\circ}18'34.76'')$, around the bulge of the host galaxy M31 (see Section~\ref{sec:observation}). We adopt 2021 June 9 (MJD 59374) as the reference epoch for the first outburst of the transient since the first X-ray detection occurred on that day at 23:44:01 UT (Section~\ref{sec:observation}). Additionally, we adopt $761 \pm 11$ kpc as the distance to the transient \citep[the same distance as the center of M31;][]{Li2021}. 

In Section \ref{sec:observation}, we present multi-epoch X-ray observations of M31 ULX-3, and our analysis of these data, including our application of the disk instability model, is described in Section \ref{sec:analysis}. We discuss our findings in Section \ref{sec:discussion} and summarize our work in Section \ref{sec:conclusion}.

\begin{figure*}[hbt!]
  \gridline{
    \fig{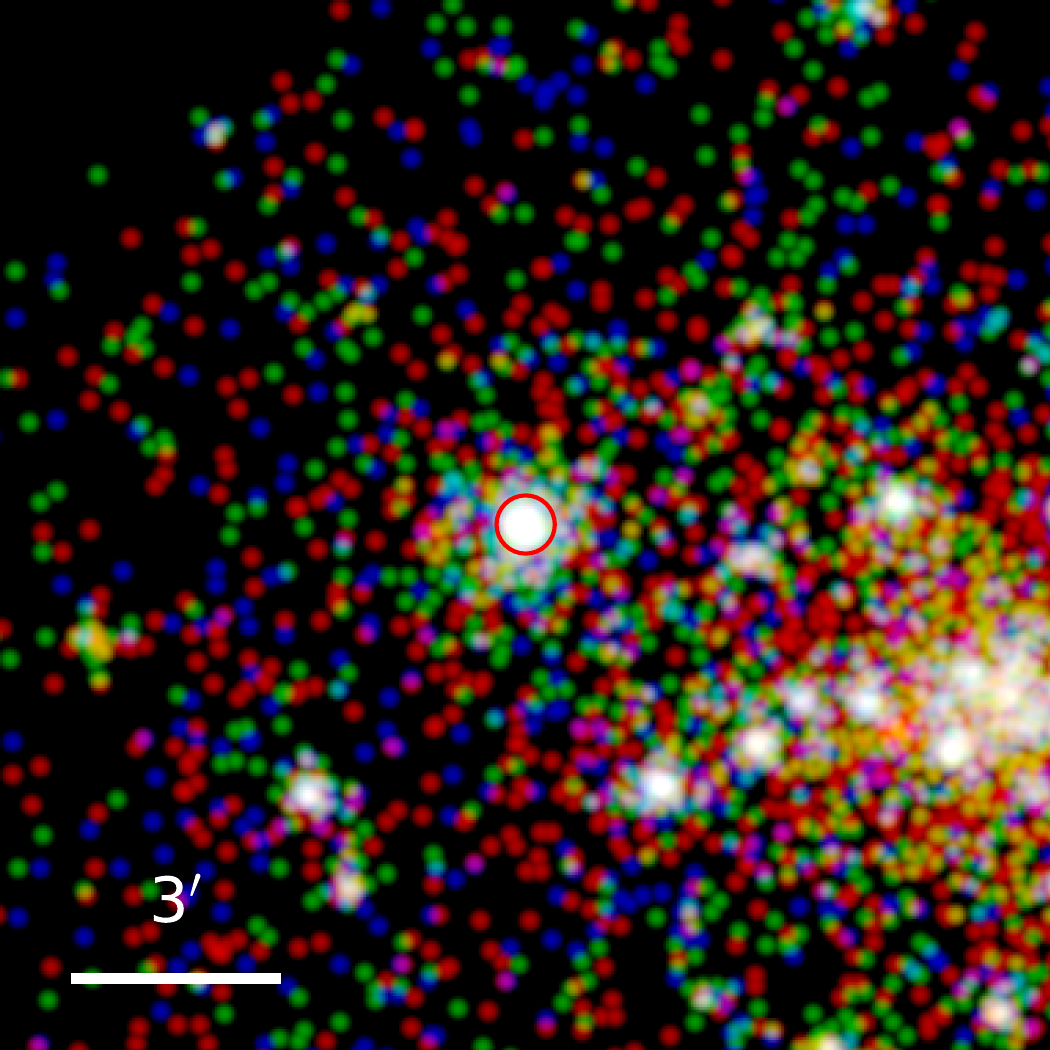}{0.25\textwidth}{}
    \fig{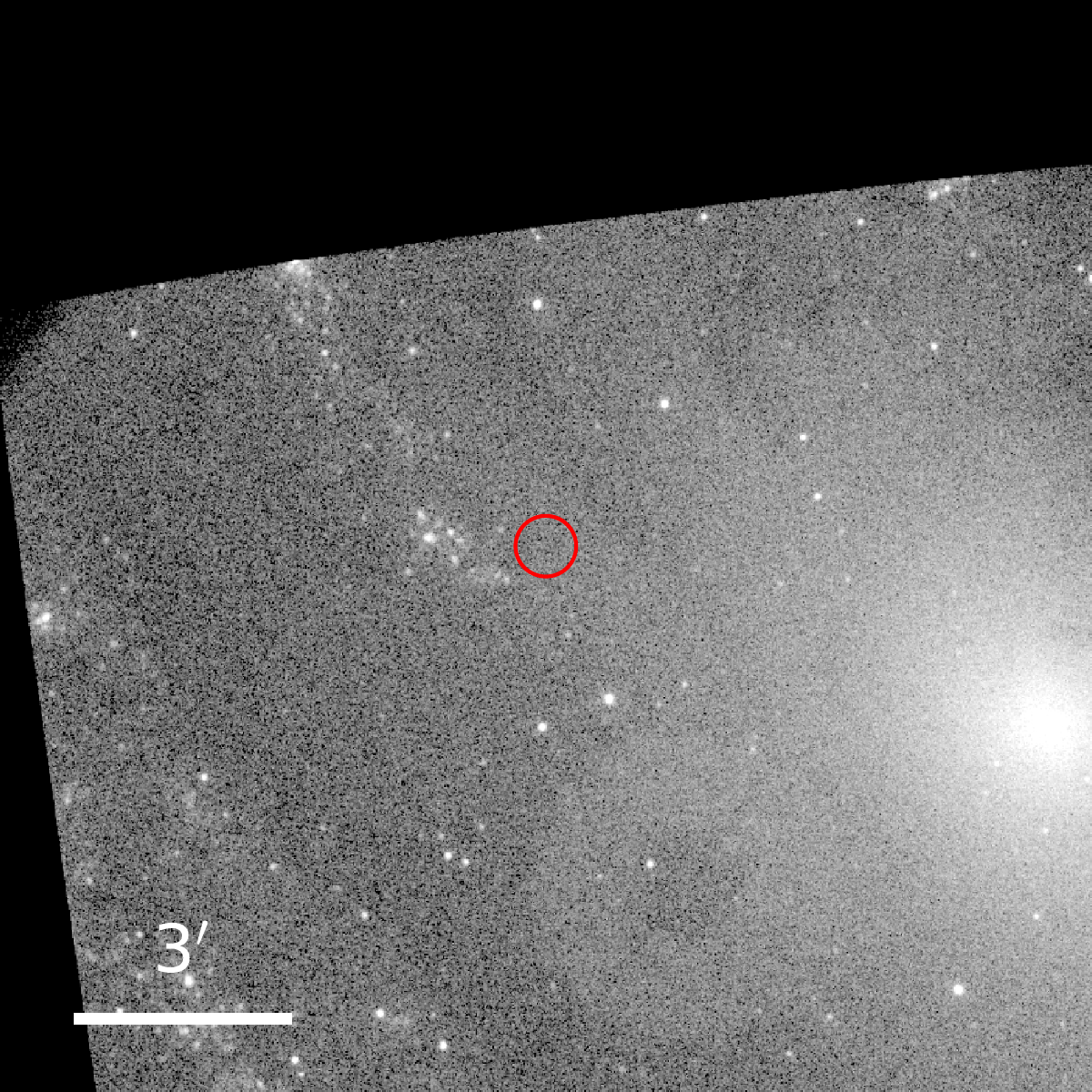}{0.25\textwidth}{}
    \fig{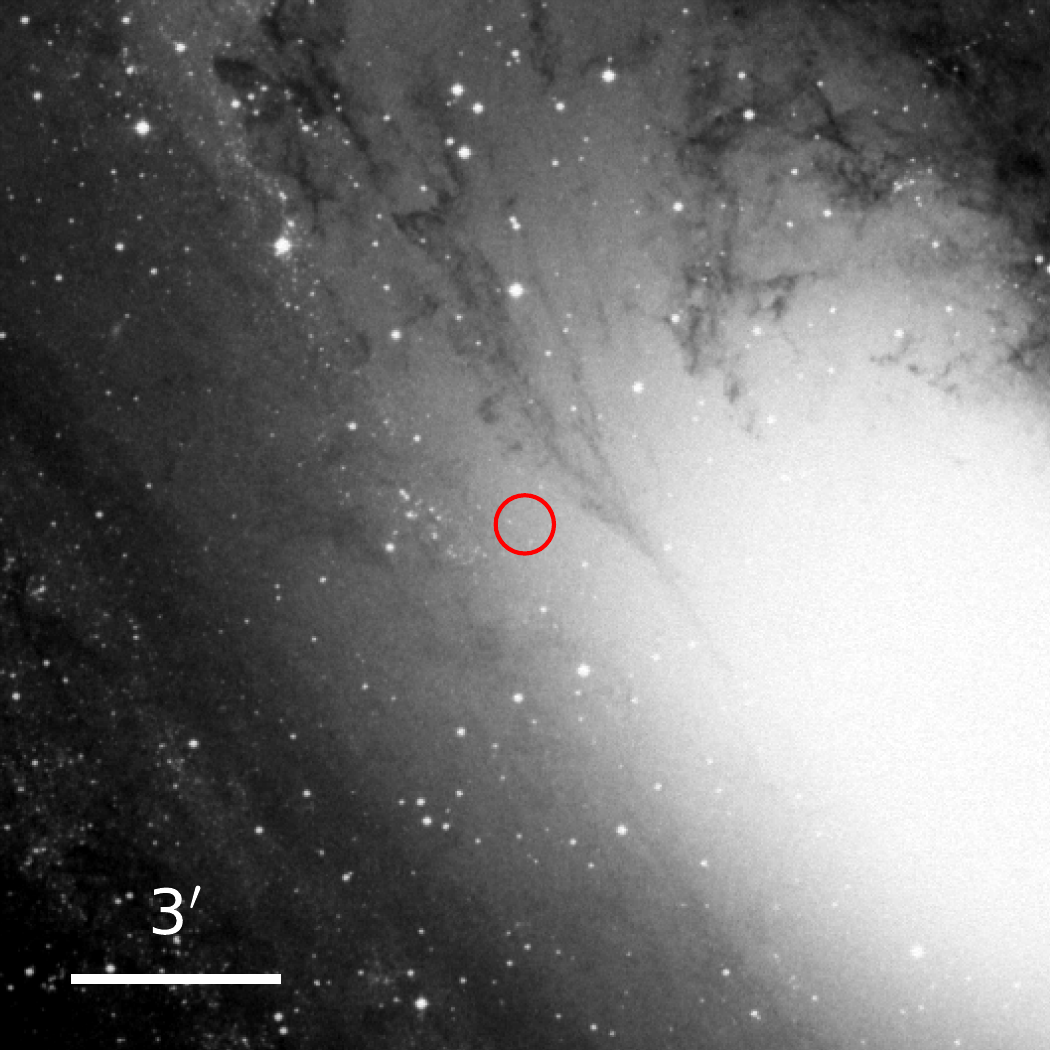}{0.25\textwidth}{}
    \fig{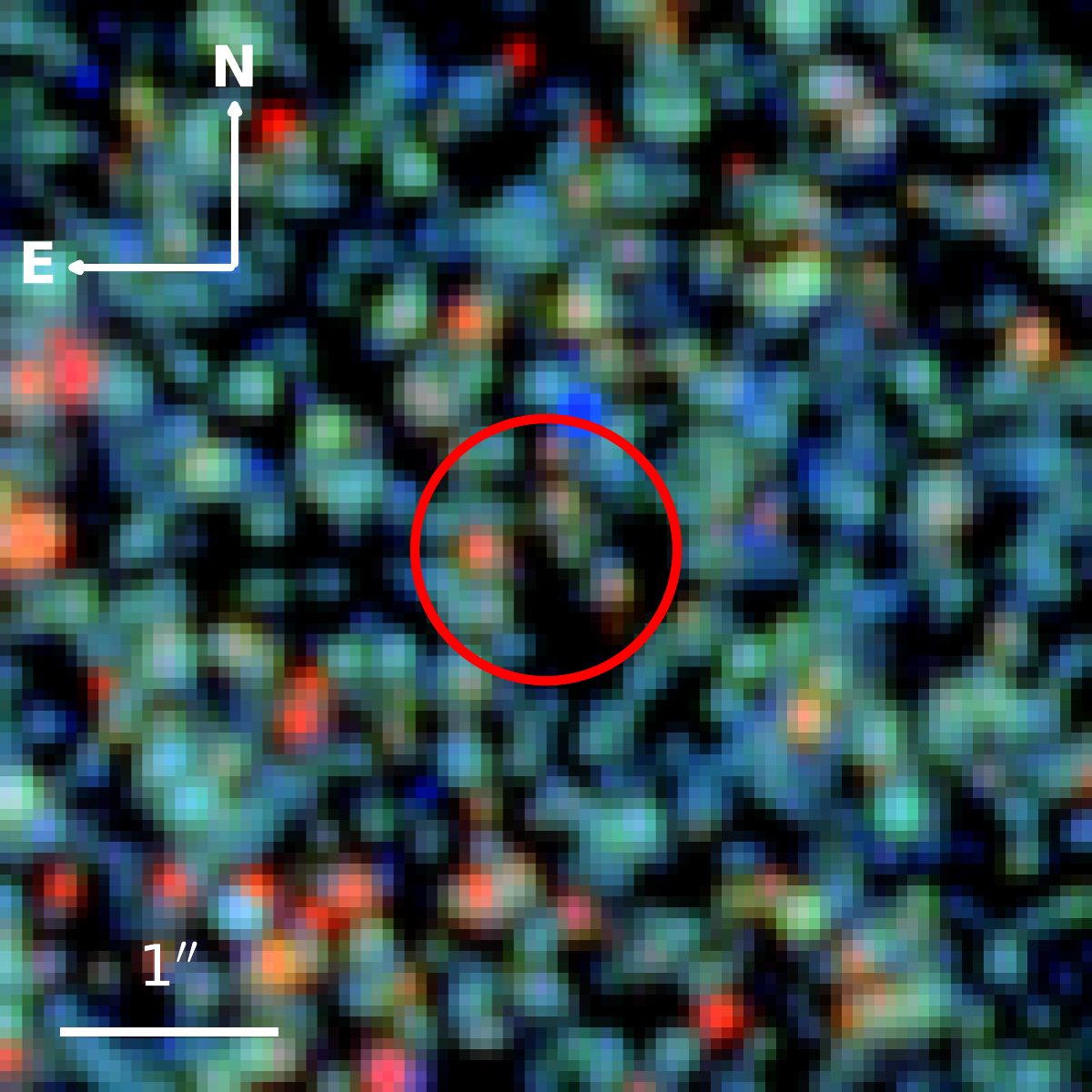}{0.25\textwidth}{}
  }
  \caption{Images of the field surrounding M31 ULX-3. From left to right: the \textit{Swift}/XRT image ($15' \times 15'$; red is $0.3-1$ keV, green is $1-2.5$ keV, and blue is $2.5-10$ keV), smoothed with a $8''$ Gaussian; the \textit{Swift}/UVOT \textit{UVW1} image ($15' \times 15'$); the DSS V-band image ($15' \times 15'$); and the \textit{HST}/ACS image ($5'' \times 5''$; red: F814W, green: F625W, and blue: F475W). The XRT, UVOT, and DSS V-band images show the corrected position using a red circle with $25''$ radius and the \textit{HST} images shows the source position with a red circle with $0.6''$ radius. A $3^{\prime}$ bar scale is shown in the \textit{Swift}/XRT, \textit{Swift}/UVOT, and DSS V-band image, while a $1^{\prime\prime}$ bar scale is shown in the \textit{HST} image. A North/East cardinal direction for all the four images is shown in the \textit{HST} image.}
  \label{fig:xrt_img}
\end{figure*}

\section{Observations} \label{sec:observation}

\subsection{Swift X-ray Telescope} \label{subsec:xrt}

M31 ULX-3, or LSXPS J004321.0+411833, was first detected serendipitiously by \textit{Swift}/XRT \citep{Burrows2005_SwiftXRT}, onboard the \textit{Neil Gehrels Swift Observatory} \citep{Gehrels2004}, on 2021 June 9 23:44:01 UT (ObsID: 00013977006; $\delta t= \text{time of observation since MJD} \ 59374= 0.99 \ \rm days$; PI: M. Middleton). An enhanced \textit{Swift}/XRT+2MASS position was obtained using the University of Leicester online tool\footnote{\url{https://www.swift.ac.uk/user_objects/}} \citep{Evans2007,Evans2009}, giving a position of $(\rm \alpha=00^h43^m21.03^s, \delta=+41^{\circ}18'33.9'')$ with an error radius of $0.9''$ (90\% confidence). The position of M31 ULX-3 appears to be around the bulge of the M31 galaxy (Figure~\ref{fig:xrt_img}). 

We also utilized the online tool to obtain the X-ray spectra. All spectra from this tool are fully calibrated and corrected and are grouped with a minimum of one count per bin using the {\tt GRPPHA} from {\tt HEASOFT v6.35} \citep{HEASoft2014}.

\subsection{Chandra X-ray Observatory} \label{subsec:cxo}

M31 ULX-3 was serendipitously detected by CXO High Resolution Camera (HRC) in two epochs: one on 2021 October 18 (ObsID: 23871; $\delta t=131.28 \ \rm days$; PI: J. Irwin) with an exposure time of 28.18 ks and the other on 2021 October 20 (ObsID: 26163; $\delta t=133.97 \ \rm days$; PI: J. Irwin) with an exposure time of 25.18 ks.

We reduced both observations using \texttt{CIAO v4.18} \citep{CIAO2006} and the corresponding calibration files. As HRC data do not contain photon energy information, source counts were extracted with \texttt{specextract} using a circular source region of radius $7''$ and a circular background region of radius $50''$. The data reduction and count extraction workflow was streamlined using \texttt{XSNAP} \citep{Ferdinand2026}.

In addition, we used the CXO HRC data (ObsID: 23871) to obtain a more precise position of M31 ULX-3. Using \texttt{wavdetect} with the default parameters, we detected a list of X-ray sources in the HRC image and crossmatched them with the Gaia Data Release 3 \citep[Gaia DR3;][]{GaiaDR3} source catalog using a matching radius of $1''$. We identified 27 matched CXO/Gaia source pairs, yielding mean astrometric offsets of $\Delta \alpha =-0.40''$ and $\Delta \delta =0.07''$. Applying these corrections, we obtained a refined position of $(\rm \alpha=0^h43^m20.93^s, \delta=41^{\circ}18'34.76'')$, with a mean residual offset of $0.6''$ between the corrected CXO and Gaia source positions. We adopt this refined position as the source position and take the mean residual offset as its positional uncertainty.

\subsection{Other X-ray Observations}

No \textit{XMM--Newton} \citep[XMM;][]{Jansen2001_XMM} or \textit{Nuclear Spectroscopic Telescope Array} \citep[NuSTAR;][]{Harrison_2013} observations covering the source were found after 2020. We queried the deepest and the most recent archival XMM observations  and estimated the upper limits around the M31 ULX-3 region from the XMM-Newton Science Archive Website\footnote{\url{https://nxsa.esac.esa.int/nxsa-web/\#search}}. The deepest observation was obtained on 2000 December 28 (ObsID: 0112570601; $t_{\rm exposure} = 13.31 \ \rm ks$), yielding a $2\sigma$ upper limit of $2.22\times10^{-3}\ {\rm counts\ s^{-1}}$, corresponding to an observed $0.2-10$ keV flux and luminosity limit of $6.97\times10^{-15}\ {\rm erg \ cm^{-2}\ s^{-1}}$ and $4.82 \times 10^{35} \ {\rm erg \ s^{-1}}$, around four order of magnitudes fainter than the peak luminosity. The most recent observation was obtained on 2016 February 9 (ObsID: 0764030401; $t_{\rm exposure} = 13.0 \ \rm ks$), yielding a $2\sigma$ upper limit of $9.27\times10^{-3}\ {\rm counts\ s^{-1}}$, corresponding to $2.79\times10^{-14}\ {\rm erg \ cm^{-2}\ s^{-1}}$ and $1.93 \times 10^{36} \ {\rm erg \ s^{-1}}$, around three order of magnitudes fainter than the peak luminosity.

We also reduced a NuSTAR observation on 2015 August 1 (ObsID: 50110001004) using NuSTAR Data Analysis Software (\texttt{NuSTARDAS}), and the NuSTAR Calibration Database (CALDB) version 20250630. We used a circular source region with a radius of $25''$ and a nearby circular background region with a radius of $125''$. No significant source emission was detected. Using a $3\sigma$ upper-limit calculation, we obtained a count-rate limit of $\sim2.0\times10^{-3}\ {\rm counts \ s^{-1}}$. Assuming the same flux-to-count conversion factor derived from the absorbed \texttt{powerlaw} model in Section~\ref{subsec:spectral_modeling}, this corresponds to an unabsorbed $0.3-10$ keV flux and luminosity limit of $\sim2.1\times10^{-13}\ {\rm erg \ cm^{-2} \ s^{-1}}$ and $\sim1.4 \times 10^{37} \ {\rm erg \ s^{-1}}$, around two order of magnitudes fainter than the peak luminosity.

\subsection{Swift Ultraviolet/Optical Telescope}

The M31 ULX-3 region was observed by \textit{Swift Ultraviolet/Optical Telescope} (\textit{Swift}/UVOT) at around the same time as \textit{Swift}/XRT. However, M31 ULX-3 was not detected in these observations (see Figure~\ref{fig:xrt_img}). We performed \texttt{uvotsource} on the \textit{Swift}/UVOT images to obtain ultraviolet (UV) upper limits of the source. We used a circular source region with $3''$ radius around the corrected position and derived a $3\sigma$ AB magnitude upper limits of $UVW1>20.66$, $UVW2>21.54$, $UVM2>21.28$, and $U>19.88$ taken from the observation obtained at $\delta t = 9.22$ days (ObsID: 00032159019; PI: Swift), when the X-ray source was luminous.

\subsection{Optical Observations}

We obtained forced photometry at the corrected X-ray position from both the Asteroid Terrestrial-impact Last Alert System\footnote{\url{https://fallingstar-data.com/forcedphot/}} \citep[ATLAS;][]{Tonry2018_ATLAS, Smith2020_ATLAS} and the Zwicky Transient Facility \citep[ZTF;][]{Bellm2019_ZTF, Masci2023_ZFPS}. We retained only unflagged measurements and found no detections above the $3\sigma$. We queried the Zwicky Transient Facility ZTF and Gaia DR3 (see Section~\ref{subsec:origins_M31}) source catalog for optical counterparts near the corrected X-ray position and found no cataloged source within $0.6''$ of the corrected X-ray position.

Archival observations show that the region around M31 ULX-3 was observed with the \textit{Hubble Space Telescope} (\textit{HST}) under programs GO12109 (PI: J. Dalcanton) and GO13710 (PI: P. Rosenfield). In this work, these \textit{HST} observations are used to produce the optical image shown in Figure~\ref{fig:xrt_img}. To identify potential optical counterparts, we instead used the \textit{Hubble Source Catalog} \citep[HSC;][]{HSC}. We identified 10 HSC sources within $0.6^{\prime\prime}$ of the corrected X-ray position of M31 ULX-3. Of these, only four have valid photometric measurements in all three filters considered here, F475W, F625W, and F814W, and we consider only these four sources in the following analysis (see Section~\ref{subsec:HST_counterparts}). The region is also covered by the Digitized Sky Survey \citep[DSS;][]{2020ipac.data.I441S}, for which we use the $V$-band image in Figure~\ref{fig:xrt_img}. 

\section{Analysis} \label{sec:analysis}

\subsection{X-ray Spectral Modeling} \label{subsec:spectral_modeling}

\begin{figure*}[tb]
    \gridline{
        \fig{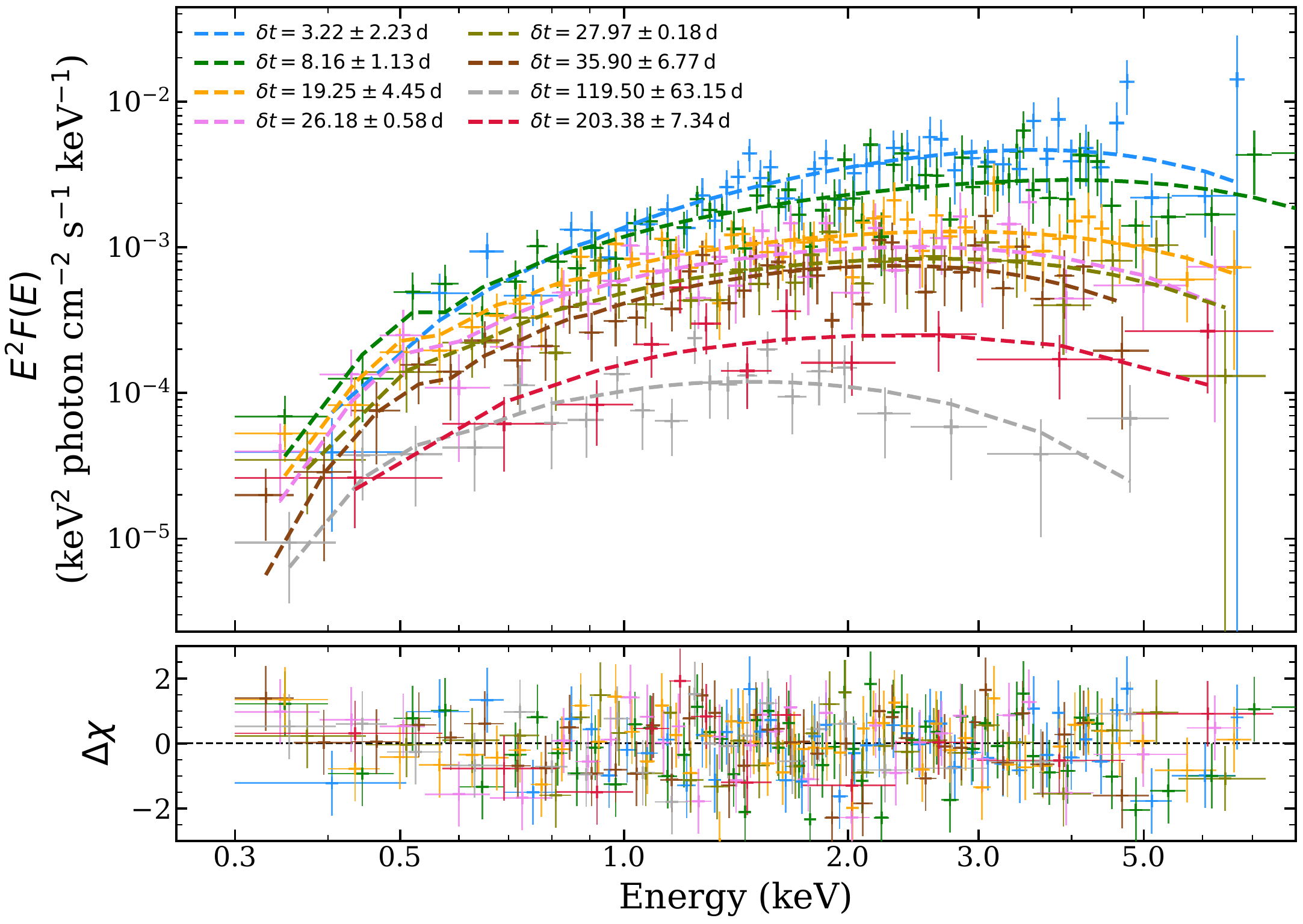}{0.75\textwidth}{}
    }
    \caption{Plot of M31 ULX-3 unfolded spectra (above) and $\Delta \chi$ (bottom) for all 8 \textit{Swift}/XRT epoch groups. A best-fit {\tt diskpbb} model is represented in a dashed line for each corresponding epoch group.}
    \label{fig:eeufspec}
\end{figure*}

\begin{figure*}[htb!]
  \gridline{
    \fig{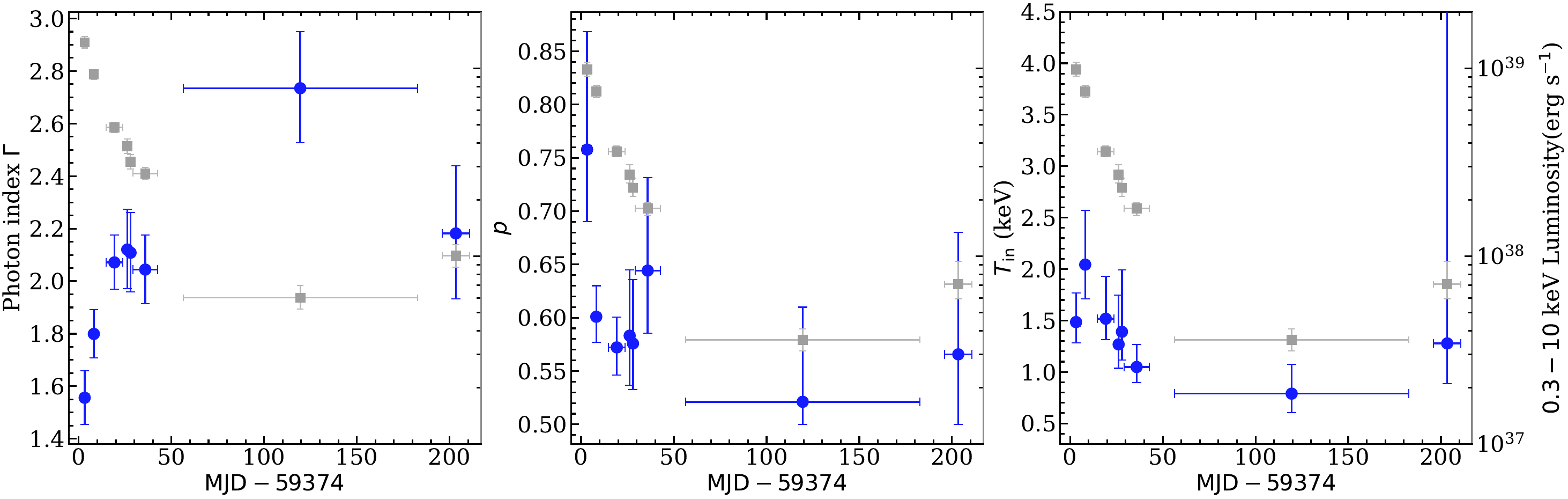}{\textwidth}{}
  }
  
  \caption{Plots of spectral evolution from \texttt{powerlaw} and \texttt{diskpbb} model (shown in blue circles) with their corresponding unabsorbed $0.3- 10$ keV luminosities (shown in gray squares) for illustration purposes. \textit{Left:} Spectral evolution of $\Gamma$.  \textit{Middle:} Spectral evolution of $p$. \textit{Right:} Spectral evolution of $T_{\rm in}$. Errors are within $1\sigma$ confidence level (corresponding to $\Delta C=1.0$). }
  \label{fig:spectral_evol}
\end{figure*}

\begin{figure*}[htb!]
  \gridline{
    \fig{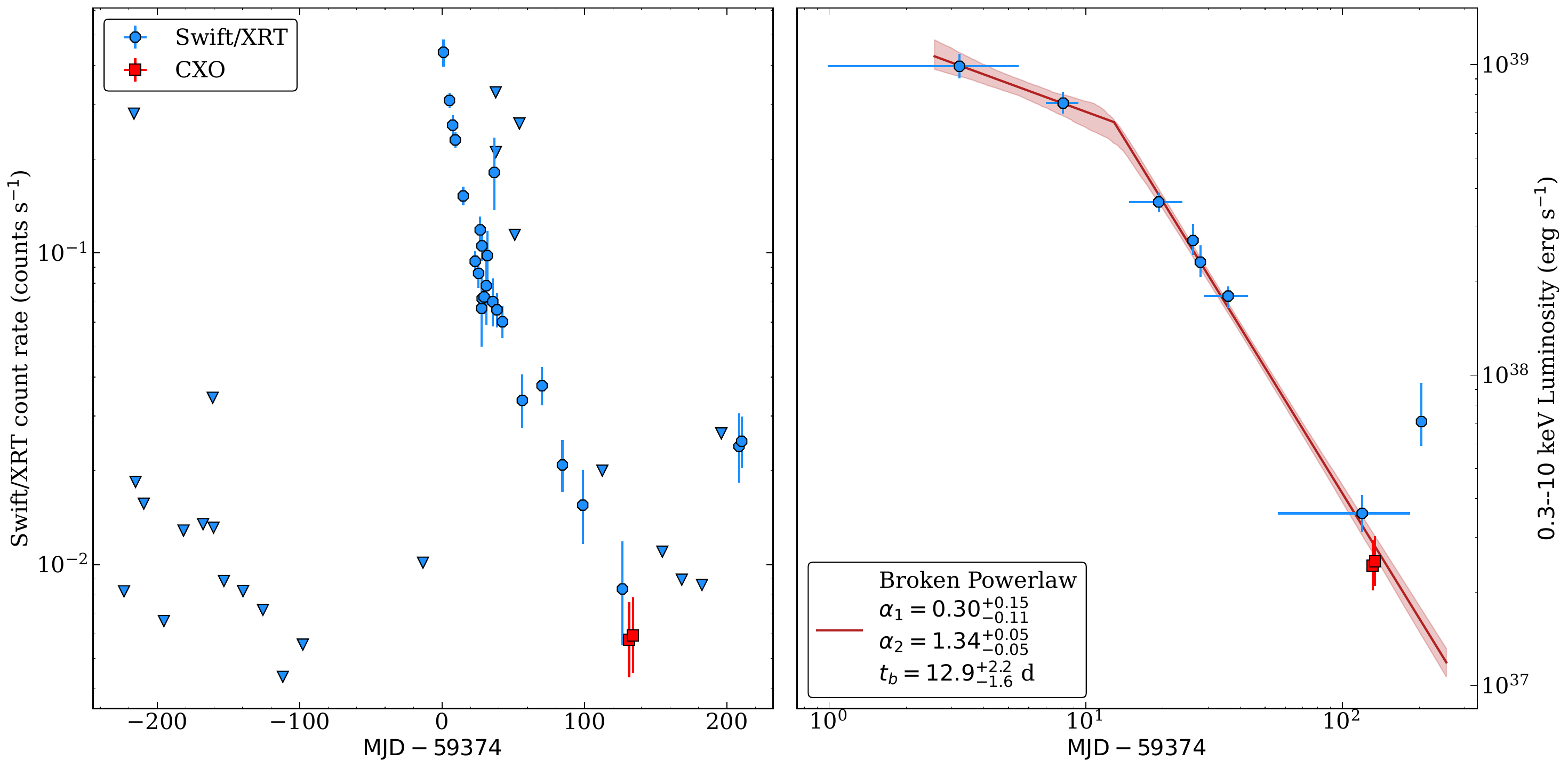}{0.95\textwidth}{}
  }
  
  \caption{\textit{Left:} Plot of count rate light curve of M31 ULX-3 from \textit{Swift}/XRT (blue circles) and CXO (red squares) observations. Blue triangles indicate \textit{Swift}/XRT upper limits. The brightening from the last nondetection to the peak count rate is $\sim 15$ days. \textit{Right:} Plot of unabsorbed $0.3 - 10$ keV X-ray luminosity light curve of M31 ULX-3 derived from the \texttt{diskpbb} model based on the epoch groups in Table~\ref{tab:fit_params}, fitted with a broken-powerlaw. The broken-powerlaw model $(L_X \propto t^{-\alpha_1} \ \text{at} \ t \leq t_b \ \ \text{and} \ \ L_X \propto t^{-\alpha_2} \ \text{at} \ t > t_b)$ is found to have $\alpha_1=0.30^{+0.15}_{-0.11}$, $\alpha_2=1.34^{+0.05}_{-0.05}$, and $t_b=12.9^{+2.2}_{-1.6} \ \rm days$. The light curve shows transient-like evolution up to $\delta t \sim 200$ days, with a late-time re-brightening at $\delta t \gtrsim 200$ days. However, its nature remains uncertain due to limited coverage.}
  \label{fig:lightcurve}
\end{figure*}

We modeled the X-ray spectra using two models: an absorbed powerlaw (hereafter \texttt{powerlaw}) and an absorbed multicolor $p$-free blackbody disk (a multiple blackbody disk model where local disk temperature $T(r) \propto r^{-p}$, where $p$ is a free parameter; hereafter \texttt{diskpbb}) model. The \texttt{powerlaw} model was used as a simple phenomenological description of the source spectra, whereas the \texttt{diskpbb} model was used to test whether the observed spectral shape could instead be explained by thermal emission from an accretion disk. In particular, we used the \texttt{tbabs*ztbabs*cflux*powerlaw} model for \texttt{powerlaw} and the \texttt{tbabs*ztbabs*cflux*diskpbb} model for \texttt{diskpbb}. The Galactic absorption column was fixed at $\rm N_{\rm H,MW}= 1.83 \times 10^{21} \ \rm cm^{-2}$ in all fits \citep{HI4PI_Collab2016}. Additionally, we adopt the solar abundances from \cite{Wilms2000}, i.e. {\tt abund wilm}, within {\tt PyXspec}.

Each observation was initially fitted individually. However, several observations had low source counts or were not significantly detected, resulting in poorly constrained spectral parameters. We therefore jointly fitted all observations after MJD 59374 with select binned epoch groups listed in Table~\ref{tab:fit_params} to improve the parameter constraints. For both \texttt{powerlaw} and \texttt{diskpbb} model joint fits, we assumed a constant intrinsic absorption, i.e. tying $\rm N_{H, int}$, froze the redshift to be $z=0$ as the source is in the M31 direction, and set the $E_{\rm min}=0.3 \ \rm  keV$ and $E_{\rm max}=10.0 \ \rm keV$ to get the unabsorbed $0.3-10 \ \rm keV$ flux. The \texttt{powerlaw} model yielded $\text{C-stat}=1244.29$ and degree of freedoms (DoFs) $=1444$ with a tied $\rm N_{H, int} = 1.05^{+0.30}_{-0.28} \times 10^{21} \ \rm cm^{-2}$. The \texttt{diskpbb} model yielded a slightly better fit, i.e. $\text{C-stat}=1216.98$ and $\text{DoFs}=1436$ $(\Delta \text{C-stat}= 27\text{ for one more free parameter})$, with a tied $\rm N_{H, int} = 1.22 \times 10^{16} \ \rm cm^{-2}$ and a $3\sigma$ upper limits of $\sim 7.7 \times 10^{20} \ \rm cm^{-2}$. The adopted joint fitting epoch groups and resulting best-fitting parameters are listed in Table~\ref{tab:fit_params}. 

The unfolded spectra and spectral data, alongside their residuals and the best-fit {\tt diskpbb} model, of each epoch group is shown in Figure~\ref{fig:eeufspec} and Figure~\ref{fig:spectra_data}, respectively. Although the spectra appear to be somewhat degenerate, the residuals indicate that the \texttt{diskpbb} model provides a good representation of the underlying spectral continuum.

The spectral evolution of the the photon index, $\Gamma$, the $p$ value, and the inner disk temperature, $T_{\rm in}$, are shown in Figure~\ref{fig:spectral_evol}. The photon index, $\Gamma$, appears to be evolving from hard to a softer state until $\sim 115$ days, and goes back to a harder state in the last epoch group; the average $\Gamma$ is $\sim2.1$ with a standard deviation of approximately $0.3$. $p$ appears to be decreasing from $\sim0.75$ to $\sim0.5$, with an average value of $\sim0.6$ and a standard deviation of $\sim0.1$. The inner disk temperature, $T_{\rm in}$, shows a tentative decline from the second epoch group of $\sim 2.0$ keV to $\sim 115$ days of $\sim 0.8$ keV. Overall, these changes indicate that the spectral shape and inferred accretion-disk properties of M31 ULX-3 evolve as the source fades.

We then converted the fitted fluxes to luminosities for each model. As the CXO HRC data does not have significant energy resolution, we estimated their luminosity using the nearest epoch group spectral parameters, i.e. the last epoch group. By comparing the total counts in each epoch group, we found an median flux-to-count conversion factor of $9.05^{+0.54}_{-1.20} \times 10^{-11} \ \rm erg \ s^{-1} \ count^{-1}$ and $6.13^{+0.61}_{-0.56} \times 10^{-11} \ \rm erg \ s^{-1} \ count^{-1}$ for the \texttt{powerlaw} and \texttt{diskpbb} model, respectively. We used this conversion factor to get the CXO HRC data in a \textit{Swift}/XRT count rate. Moreover, we fitted the luminosity with a broken-powerlaw model $(L_X \propto t^{-\alpha_1} \ \text{at} \ t \leq t_b \ \ \text{and} \ \ L_X \propto t^{-\alpha_2} \ \text{at} \ t > t_b)$, producing a fit of $\chi^2/\text{DoF}=28.15/8$ and best-fitting parameters $\alpha_1=0.30^{+0.15}_{-0.11}$, $\alpha_2=1.34^{+0.05}_{-0.05}$, and $t_b=12.9^{+2.2}_{-1.6} \ \rm days$. 

The \textit{Swift}/XRT count rate and luminosity derived from the \texttt{powerlaw} model light curve are shown in Figure~\ref{fig:lightcurve}. The light curve of M31 ULX-3 generally exhibits transient-like behavior up to $\delta t \sim 200$ days, followed by a re-brightening at $\delta t \gtrsim 200$ days. This late-time behavior could reflect a weak re-flare or renewed accretion, which can sometimes be seen in Galactic LMXBs after major outbursts \citep[e.g., V404 Cyg and MAXI J1820][]{Sanchez2018, Thomas2022}. Another possibility is that the source spectrum evolves as the flux decreases. However, this interpretation remains speculative due to the lack of observations beyond $\delta t \sim 200$ days.

\subsection{Disk-instability Model} \label{subsec:dim}

\begin{deluxetable}{ccccccccccc}
\tabletypesize{\small}
\tablewidth{0pt}

\tablecaption{Fits of the Disk-instability Model
\label{tab:dim_lightcurve_fits}}

\tablehead{
\colhead{} &
\colhead{} &
\multicolumn{4}{c}{\texttt{powerlaw} Model} &
\colhead{} &
\multicolumn{4}{c}{\texttt{diskpbb} Model}
\\
\cline{3-6}
\cline{8-11}
\colhead{Source Name} &
\colhead{$M_1$} &
\colhead{$t_0$} &
\colhead{$\dot{M}_{\max}/\dot{M}_{\rm Edd}$} &
\colhead{$\alpha$} &
\colhead{$\chi^2/{\rm DOF}$} &
\colhead{} &
\colhead{$t_0$} &
\colhead{$\dot{M}_{\max}/\dot{M}_{\rm Edd}$} &
\colhead{$\alpha$} &
\colhead{$\chi^2/{\rm DOF}$}
\\[-7pt]
\colhead{} &
\colhead{} &
\colhead{(days)} &
\colhead{} &
\colhead{} &
\colhead{} &
\colhead{} &
\colhead{(days)} &
\colhead{} &
\colhead{} &
\colhead{}
}
\startdata
LSXPS J004321.0+411833
& 1.4 & 49.6 & 11.46 & $<0.20$ & 26.94/7 &
& 41.8 & 8.98 & $<0.20$ & 40.36/7
\\[1pt]
(M31 ULX-3) & 10 & 60.3 & 1.04 & $<0.60$ & 29.07/7 &
& 57.0 & 0.78 & $<0.60$ & 21.97/7
\\
\enddata
\tablecomments{Fit excludes the final \textit{Swift}/XRT observation as the model assumes a smooth decline.}

\end{deluxetable}

\begin{figure*}[htb!]
  \gridline{
    \fig{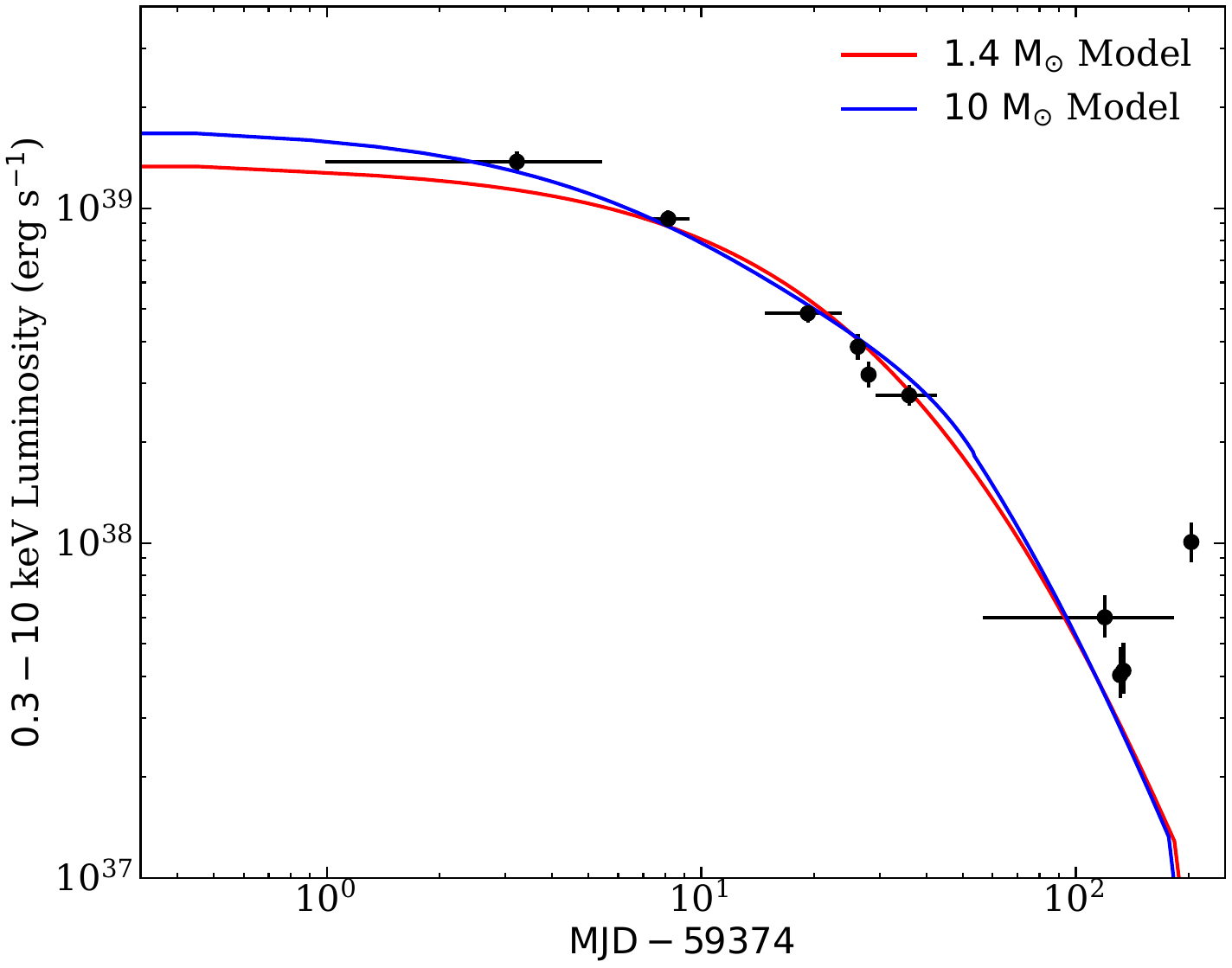}{0.5\textwidth}{}
    \fig{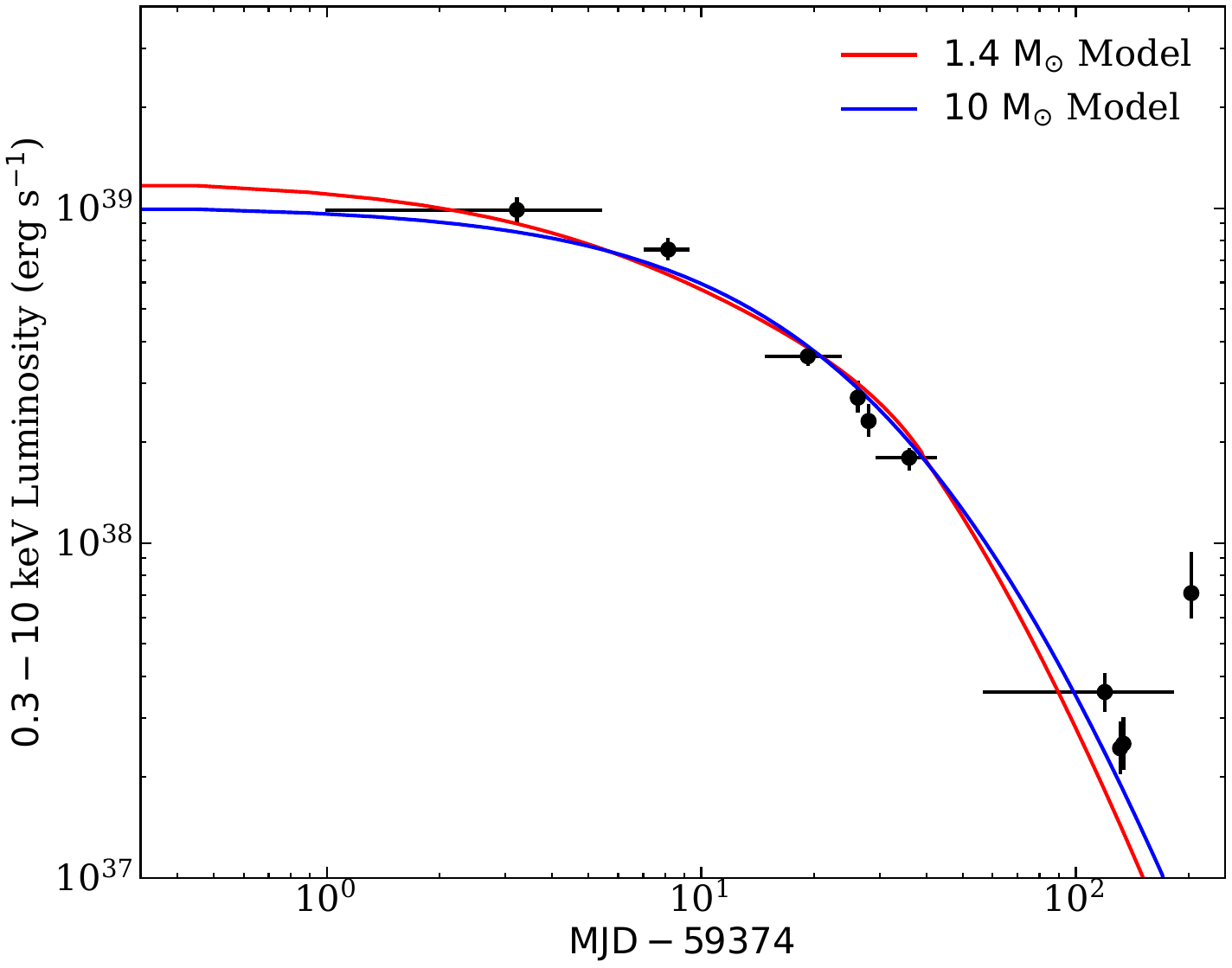}{0.5\textwidth}{}
  }
  
  \caption{Plot of fitted unabsorbed $0.3-10$ keV X-ray luminosity light curve, derived from the \texttt{powerlaw} model (left) and the \texttt{diskpbb} model (right), with the disk-instability model. Black points indicate the unabsorbed $0.3-10$ keV X-ray luminosity. The red and blue lines show the best-fitting disk-instability models for assumed compact-object masses of $1.4\,M_{\odot}$ and $10\,M_{\odot}$, respectively. Because the disk-instability model predicts only a smooth decline, the final \textit{Swift}/XRT observation was excluded from the fits.}
  \label{fig:dim_lc_plot}
\end{figure*}

To investigate the physical mechanism responsible for the outburst, we modeled the X-ray light curve using the disk-instability framework of
\citet{Hameury2020}. Although it is initially used to explain dwarf nov\ae\ and other X-ray transients, this model has been tested and proven to be able to reproduce transient ULX light curves in previous studies \citep[e.g.,][]{Brightman2020, Brightman2023}. According to this model, the outburst is triggered by a thermal--viscous instability that drives the accretion disk into a hot, ionized state. The mass-accretion rate during the initial decay is
approximated by

\begin{equation}
    \dot{M}(t) = \dot{M}_{\max} \left(1+\frac{t}{t_0}\right)^{-10/3},
    \label{eq:dim_mdot}
\end{equation}
where $\dot{M}_{\max}$ is the initial or maximum mass-accretion rate and $t_{0}$ is the characteristic decay timescale. At $t=t_0$, the
mass-accretion rate has decreased to approximately one tenth of its initial value. The characteristic timescale, $t_0$ is related to the physical properties of the accretion disk by

\begin{equation}
    t_0 = 3.19 \,\alpha_{0.2}^{-4/5} M_1^{1/4} r_{12}^{5/4} \dot{M}_{\max,19}^{-3/10}  \ {\rm yr},
    \label{eq:dim_t0}
\end{equation}
where $\alpha_{0.2}=\alpha/0.2$ is the Shakura--Sunyaev viscosity parameter \citep{Shakura1973}, $M_1$ is the accretor mass in units of $M_{\odot}$, $r_{12}$ is the characteristic disk radius in $10^{12} \ \rm cm$, and
$\dot{M}_{\max,19}$ is $\dot{M}_{\max}$ in $10^{19} \ \rm g \ s^{-1}$. Thus, for an assumed accretor mass, viscosity, and size, the observed decay timescale can be used to constrain the the peak disk accretion rate.

The initial viscous-decay phase continues until the accretion rate drops below the critical value required to keep the outer disk in the hot state. Its duration is approximated by

\begin{equation}
    \Delta t_1 = t_0 \left[ 1.38\, t_{0}^{-0.50} M_{1}^{0.25} \dot{M}_{19,\max}^{0.15} f_{\rm irr}^{0.15} \alpha_{0.2}^{-0.4} -1 \right],
    \label{eq:dim_dt1}
\end{equation}
where $f_{\rm irr}$ parameterizes the effects of irradiation to the outer disk. Once the outer disk can no longer remain in the hot state, a cooling front propagates inward and produces a more rapid final decline. The duration of this phase is

\begin{equation}
    \Delta t_2 = 0.74\, M_1^{0.37} f_{\rm irr}^{0.15} \alpha_{0.2}^{-0.8} r_{12}^{0.62} \ {\rm yr}.
    \label{eq:dim_dt2}
\end{equation}
The total modeled outburst duration is therefore $\Delta t_{\rm out} = \Delta t_{1}+\Delta t_{2}$.

For super-Eddington accretion rates, we converted the modeled mass-accretion rate into an X-ray luminosity using
\begin{equation}
    L_{\rm X} = (1+\ln \dot{m})[1+ \dot{m}^2/b] \, L_{\rm Edd}    
    \label{eq:dim_luminosity}
\end{equation}
where $\dot{m}=\dot{M}/\dot{M}_{\rm Edd}$ and $b$ is a constant related to the beaming parameter \citep[see][]{King2009}. In this work, we adopt $b=73$, following \citet{King2009}. This modified prescription provides a smooth transition where beaming is negligble, as explained in \citet{Hameury2020}.

We fitted the luminosity light curves of the \textit{Swift}/XRT and CXO observations calculated from the \texttt{powerlaw} and \texttt{diskpbb} spectral models separately (Section~\ref{subsec:spectral_modeling}). As the model only predicts a smooth decline, we excluded the last \textit{Swift}/XRT observation from the fits. For each light curve, we considered compact-object masses of $M_{1}=1.4\,M_{\odot}$ and $10\,M_{\odot}$, representing neutron star and stellar-mass black hole accretors, respectively. The fitted parameters were the peak accretion-rate, $\dot{M}_{\max}$, the characteristic decay timescale, $t_{0}$, and the viscosity parameter, $\alpha$. 

The resulting parameters are reported in Table~\ref{tab:dim_lightcurve_fits}, and the corresponding model light curves are shown in Figure~\ref{fig:dim_lc_plot}. Based on the fits, the disk-instability model favors the $10\,M_{\odot}$ model for luminosities derived from the {\tt diskpbb} model, whereas neither $1.4\,M_{\odot}$ nor $10\,M_{\odot}$ model is clearly preferred for luminosities derived from the {\tt powerlaw} model. Comparing to Figure~\ref{fig:lightcurve}, the fitted broken-powerlaw light curve shows a better fit than the model adopted here. Because the observations do not clearly resolve the transition to the final decay, $\alpha$ is not tightly constrained. The inferred $\alpha$ values should therefore be regarded as upper limits, and the derived $t_0$ and $\dot{M}_{\max}/\dot{M}_{\rm Edd}$ should be interpreted with this limitation in mind.

Based on the best-fit parameters, the maximum accretion rate is larger than the Eddington limit for the $M_1=1.4 \, M_{\odot}$ model and is around the Eddington limit for $M_1=10\, M_{\odot}$ model. Excluding the last \textit{Swift}/XRT observation, the fits reproduce the observed light curve adequately. Although a slight discrepancy is apparent at $\delta t \sim 100- 200\ \rm days$, this may simply reflect the broad epoch bins and the correspondingly large uncertainty in $\delta t$. Alternatively, the remaining difference may indicate that some physical processes are not included in the simple disk-instability model.

%None of the fits fully reproduces the observed light curve perfectly. The main difference appears at late times $(\delta t\sim 100-200 \ \rm days)$, where it appears to have a  re-brightening, while the model predicts a smooth decline. 
%The late-time behavior could reflect a weak re-flare or renewed accretion, which can sometimes be seen in Galactic LMXBs after major outbursts \citep[e.g., V404 Cyg and MAXI J1820][]{Sanchez2018, Thomas2022}. Another possibility is that the source spectrum evolves as the flux decreases. As the luminosities from CXO HRC observations were estimated using an assumed spectral shape, such spectral changes could affect the inferred luminosities. The remaining difference may also indicate that some physical processes are not included in the simple disk-instability model. However, this interpretation remains speculative as the fits do not reproduce the light curve well and we lack observations at $\delta t \gtrsim200 \ \rm days$. 

\section{Discussion} \label{sec:discussion}

\begin{figure*}[tb]
    \gridline{
        \fig{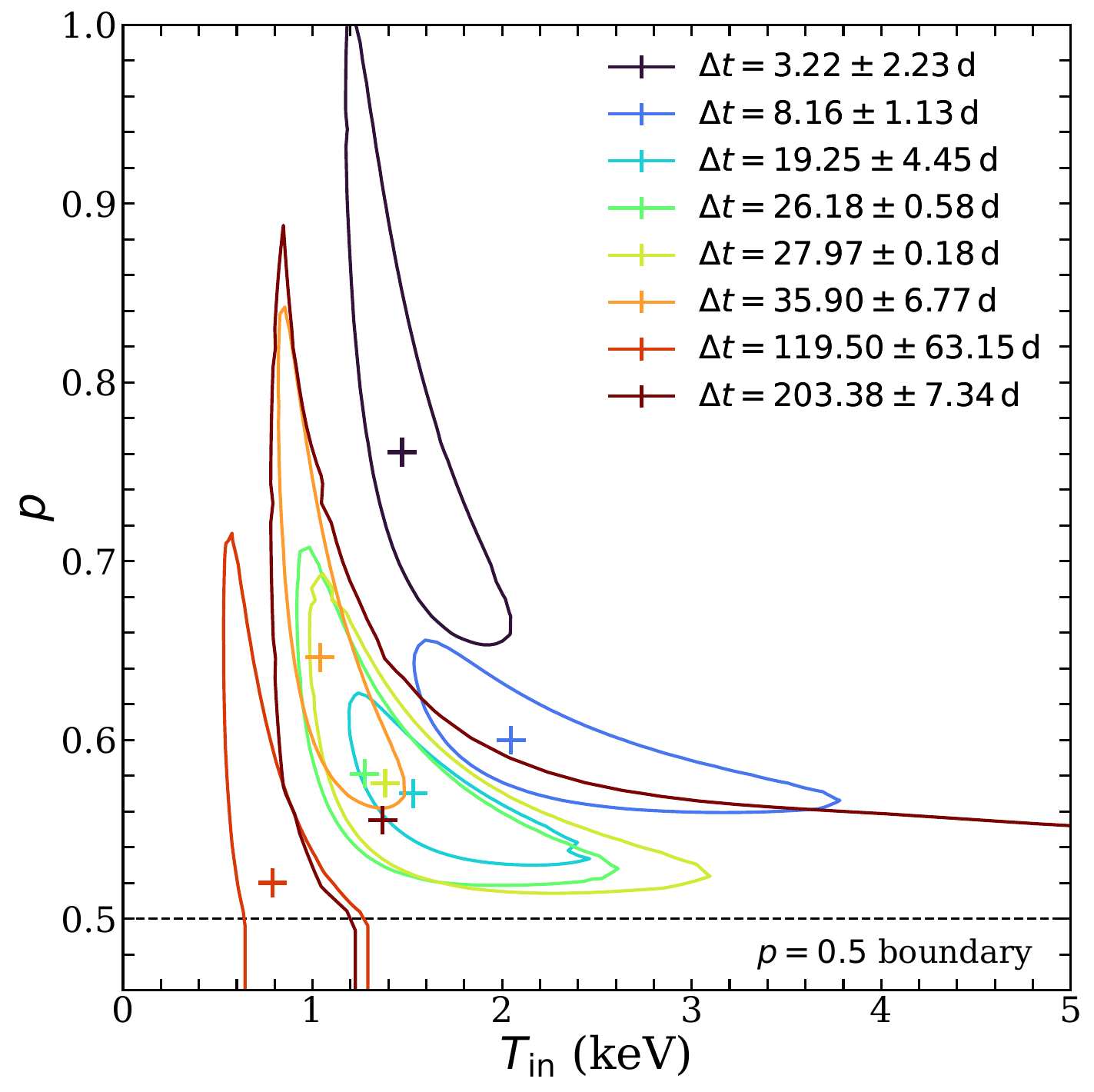}{0.5\textwidth}{}
        \fig{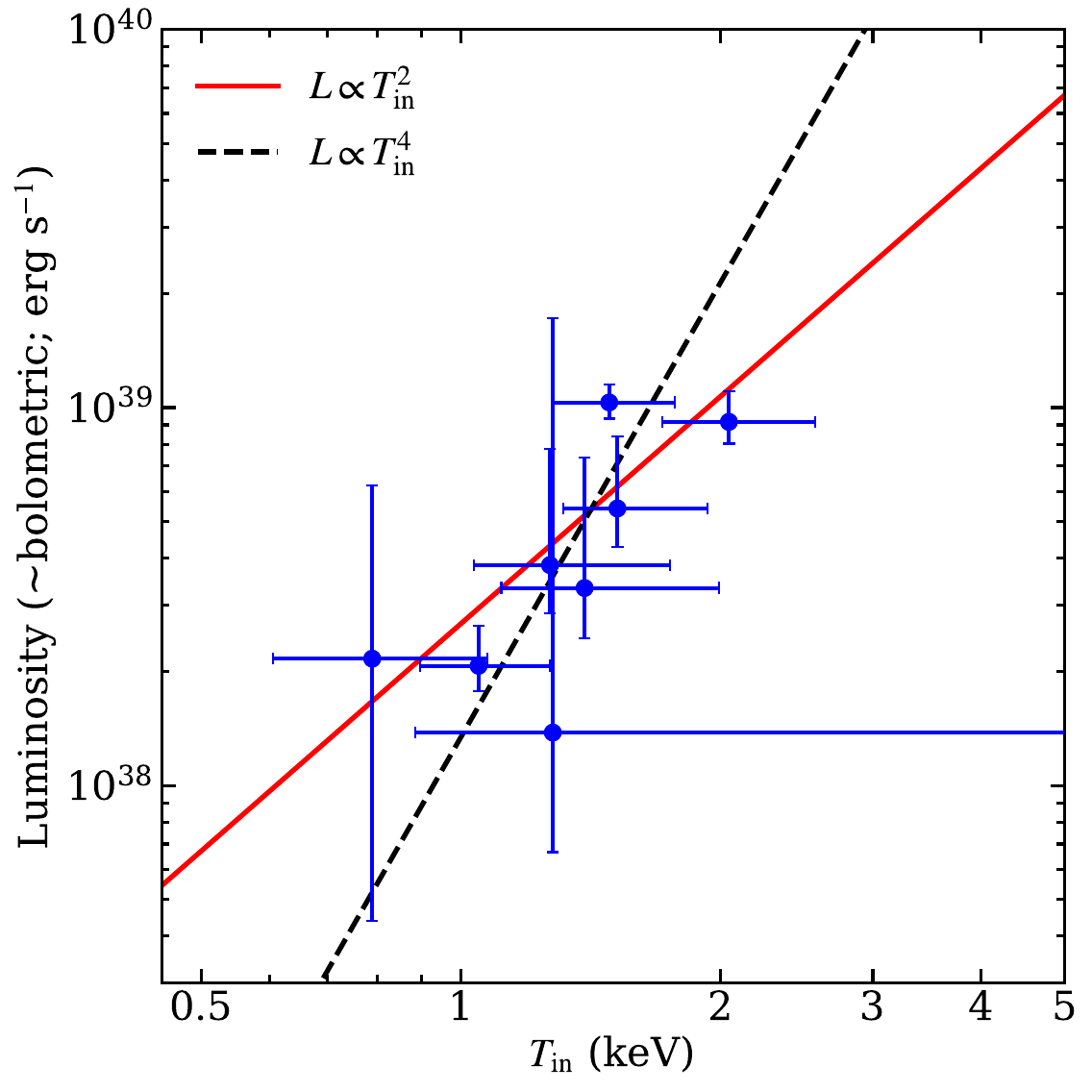}{0.5\textwidth}{}
    }
    \caption{\textit{Left:} Evolution of the spectral-fit of the \texttt{diskpbb} model parameters $p$ and $T_{\rm in}$ with respect to time since MJD 59374. The $1\sigma$ joint confidence contours (corresponding to $\Delta C=2.30$) are shown for each \textit{Swift}/XRT observation in Table~\ref{tab:fit_params}, and the best-fit values are marked by crosses. The earliest observations constrain relatively higher values of $p$, while later observations approach the lower allowed boundary of $p=0.5$. The dashed black line indicates this imposed lower limit. \textit{Right:} Unabsorbed $0.001–100$ keV \citep[$\sim$ bolometric, following][]{Walton2025} luminosity versus $T_{\rm in}$ for the {\tt diskpbb} model of each \textit{Swift}/XRT observation in Table~\ref{tab:fit_params}, indicated by blue circles. The luminosity--temperature relation seems to be broadly consistent with $L \propto T^2$ (red solid line) and $L\propto T^4$ (black dashed line).}
    \label{fig:contour_p_Tin}
\end{figure*}

\subsection{Possible Origins of the M31 ULX-3} \label{subsec:origins_M31}
% See if it's foreground or background
The nature of M31 ULX-3 is not immediately evident from its observed X-ray properties alone. While its projected location and inferred luminosity are consistent with a ULX in M31, alternative interpretations, including a foreground Galactic source or a background extragalactic object, must also be considered.

\subsubsection{A Foreground Galactic Source?}

If we assume the source is in our Milky Way Galaxy at a distance of $\sim 10$ kpc, its peak X-ray luminosity would be $\sim 10^{35} \ \rm erg \ s^{-1}$ and time above half-maximum light around $\sim20$ days.  Its position in the plane defined by peak luminosity vs time above half-maximum brightness overlaps with the parameter space exhibited by several Galactic transient populations, including classical novae, XRBs, and magnetar flares \citep{Polzin2023}. M31 ULX-3 seems unlikely to be a classical nova as they are usually accompanied by optical/UV flares \citep[e.g.][]{Page2020}, whereas no corresponding transient was identified at these wavelengths. The absence of optical or UV emission would also disfavor the possibility of high-mass XRBs or Be XRBs \citep{Fornasini2023}. It is less decisive for a fainter XRB, such as LMXBs, or other fainter X-ray transients, such as those found in \citet{Degenaar2010}, and magnetars as optical/UV counterparts of such systems may not be detectable \citep{Esposito2021_Magnetars, Wang2024}. However, the lack of a Gaia source around the corrected position provides additional evidence against a nearby Galactic stellar counterpart. A Galactic origin therefore appears to be unlikely.

\subsubsection{A Background Extragalactic Source?} 

Although M31 ULX-3 is projected onto the outer bulge of M31 (Figure~\ref{fig:xrt_img}), it could still be an unrelated background transient. If M31 ULX-3 were located at $z\sim0.1$, corresponding to a luminosity distance of $\sim460 \ \rm Mpc$, its peak X-ray luminosity would be $\sim10^{44}-10^{45} \ \rm erg \ s^{-1}$. Such luminosities and timescales are broadly consistent with those observed for tidal disruption events \citep[TDEs; see][]{Auchettl2017}. However, a background TDE at this distance would be expected to reside in a host galaxy. Known TDE host galaxies at comparable distances are typically relatively luminous, with $V$-band magnitudes of $m_V\sim17.5-21.5$ \citep[see][]{French2020}, whereas HSC sources near the corrected position of M31 ULX-3 are mostly fainter than $m_{\rm F625W}\sim23.0$ (Section~\ref{subsec:HST_counterparts}). 

The second segment of the broken-powerlaw (Figure~\ref{fig:lightcurve}) follows $L_{\rm X}\propto t^{-1.34}$, broadly similar to the canonical TDE fallback scaling of $t^{-5/3}$ \citep{Rees1988, Phinney1989}, which may describe a TDE-like behavior. However, M31 ULX-3's X-ray properties do not match with those of TDEs. Usually, TDEs have soft X-ray energy with disk temperature $\sim 0.1 \ \rm keV$ \citep[see][for a comprehensive review]{Saxton2020}, while M31 ULX-3 disk temperature is $\sim 1-2 \ \rm keV$. Additionally, although foreground extinction and crowding from the M31 bulge could obscure a faint background host galaxy and reduce the detectability of an optical/UV counterpart, no evidence for additional absorption was found in the X-rays (Section~\ref{subsec:spectral_modeling}), which argue against a background TDE source. The absence of a plausible host galaxy candidate in the deep \textit{HST} imaging and the lack of bright optical or UV flares (Figure~\ref{fig:xrt_img}) argue against M31 ULX-3 as a background extragalactic source.

\subsubsection{Association with M31?}

As M31 ULX-3 is unlikely to be a Galactic or background source, we can safely assume that M31 ULX-3 is associated with the M31 galaxy. At this distance, with $L_{X, \rm peak} \sim 10^{39} \ \rm erg \ s^{-1}$ and time above half-maximum light $\sim 10$ days, it is possible for M31 ULX-3 to be a supernova \citep{Chevalier2017, Page2020}. However, a supernova in 2021 in M31, a galaxy that is frequently monitored, would likely have been detected in other wavelengths, especially in the optical and UV. The lack of any such counterpart at these wavelengths again makes this scenario unlikely. We therefore conclude that M31 ULX-3 likely is indeed another transient ULX in M31.

\subsection{Nature of the M31 ULX-3 Accretor}

Based on the results in Section~\ref{subsec:spectral_modeling}, we found M31 ULX-3 spectra to be evolving as the luminosity fades, especially in the $p$-value from the \texttt{diskpbb} model and the photon index, $\Gamma$, from the \texttt{powerlaw} model.

The \texttt{diskpbb} fits suggest that M31 ULX-3 retains a broadened disk-like spectrum throughout its spectral evolution. As shown in Figure~\ref{fig:spectral_evol} and Table~\ref{tab:fit_params}, the $p$-value decreases from $p=0.76$ near the beginning of the outburst to close to the lower model boundary of $p=0.5$ in the final two epochs. The joint $1\sigma$ confidence contours (Figure~\ref{fig:contour_p_Tin}) show that this evolution is accompanied by a subtly decreasing inner-disk temperature and that the early-time and late-time solutions occupy different regions of the  $p-T_{\rm in}$ parameter space. For comparison, a standard, non-relativistic thin disk follows $T(r) \propto r^{-p}$ with $p=0.75$, whereas $p\simeq 0.5$ is characteristic of a flatter temperature profile expected from a slim advective disk. Taken at face value, the fits suggest that the spectrum evolves from a standard thin disk profile toward a substantially broader, slim advective disk. One must however keep in mind that the standard thin disk does not include relativistic corrections and cannot account for super-Eddington luminosities. The goodness of the fit does not imply that the physical description is correct, and other ingredients must probably be added as discussed below.

As $T_{\rm in}$ appears to slightly decrease as the source fades, we also examined the luminosity--temperature relation. Following \citet{Walton2025}, we estimated the unabsorbed luminosity over the $0.001 - 100$ keV band as an approximation to the unabsorbed bolometric luminosity and compared it with $T_{\rm in}$. Figure~\ref{fig:contour_p_Tin} shows that the observed evolution is broadly consistent with both $L\propto T^4$, as expected for blackbody emissions with a constant emitting area, and the shallower $L\propto T^2$ relation, associated with emissions from a slim, advective disk \citep{Watarai2000}.  Given the limited number of epochs and large uncertainties in the data, it is hard to robustly compute the slope of the $L - T_{\rm in}$ relationship. Nevertheless, the data do not exclude the $L\propto T^2$ relation expected for a slim disk, consistent with the lower $p$-values inferred from the \texttt{diskpbb} fits at late times.

This apparent evolution is nevertheless counterintuitive. Radial advection is normally expected to become important at the highest accretion rates, whereas the lowest $p$-values in M31 ULX-3 occur as the source fades. %It is also interesting that $T_{\rm in}$ remains broadly constant despite the changes in luminosity and $p$, whereas variations by factors of $\sim2$--$2.5$ are more commonly observed. 
A qualitatively similar, although weaker, trend was reported for the transient M31 ULX-1, whose $p$-value decreased from approximately 0.60 to 0.55 during its decline \citep{Middleton2012_ULX1}. In M31 ULX-1, however, $T_{\rm in}$ increased as $p$ decreased, in contrast to the M31 ULX-3 subtly decreasing $T_{\rm in}$ as $p$ also decrease. 

\citet{Middleton2012_ULX1} noted that the strong advection inferred at lower luminosities was inconsistent with the estimated sub-Eddington accretion rate and suggested that this discrepancy could partly reflect the limitations of the \texttt{diskpbb} model. As the model does not include relativistic spectral smearing and therefore produces an intrinsically narrower disk spectrum, the fit may compensate for the missing broadening by favoring a lower $p$. 

Another possibility is that a steep Comptonized component becomes relatively more prominent as the source fades, increasing the contribution of hard emission and altering the apparent shape of the disk continuum \citep[e.g.,][]{Gladstone2009, Middleton2011, Middleton2012_ULX1, Sutton2013}. If the spectrum is instead fitted with a single \texttt{diskpbb} component, such an additional hard component could bias the inferred \(p\) value. M31 ULX-1 is a good example of this as it favors a two-component spectrum consisting of an advective disk and an additional Comptonized component \citep{Middleton2012_ULX1}. Given the limited sensitivity of the current X-ray spectra, however, we cannot determine whether such a component is present or whether it is responsible for the apparent evolution in $p$.

Alternatively, M31 ULX-3 may evolve from a standard thin disk to a radiation-pressure-dominated inner disk once above $\sim0.3L_{\rm Edd}$. The inner disk may produce overlapping thermal components and an advective spectrum, while the presence of an optically thick wind may further modify the continuum. The declining $p$-value could reflect changing contributions from the inner disk and wind photosphere, or degeneracy from fitting this composite spectrum with a single-component \texttt{diskpbb} model. Nonetheless, these interpretations remain tentative and there is no reason to believe that the emitted spectrum is a pure blackbody disk. This is due to spectral distortion that the color temperature from the spectrum is often different than the effective temperature for X-ray bursts from neutron stars \citep[see][]{Rybicki1979, London1986, Suleimanov2011}. 

The \texttt{powerlaw} fits may provide additional support for an evolving wind-dominated super-Eddington accretion. Figure~\ref{fig:spectral_evol} shows a clear spectral softening as its X-ray luminosity and $p$-values decline, with the photon index increasing from $\sim 1.6$ at early times to $\sim2.7$ at $\sim115$ days. Since the intrinsic absorbing column was tied across the observations at $\rm N_{H, int}\sim 10^{21} \ \mathrm{cm^{-2}}$, this trend is less likely to be driven solely by degeneracy between absorption and the continuum slope. However, such a scenario can arise in super-Eddington accretion as it may drive an increasingly optically thick wind, which can scatter and reprocess the hard inner emission and produce a softer spectrum and a lower observed X-ray luminosity \citep{Sutton2013,Middleton2015,Pinto2017,Kaaret2017}. Taken together, the evolution of $p$ and $\Gamma$ is generally consistent with a wind-dominated super-Eddington flow, although constraints on the accretion rate are still unknown.

Regardless, M31 ULX-3 appears to have an evolving radiative processes as the source fades and has been suggested by both the \texttt{diskpbb} and the \texttt{powerlaw} model (Section~\ref{subsec:spectral_modeling}).  Further broadband observations and spectral modeling, particularly if M31 ULX-3 undergoes another outburst, will be required to distinguish between these possibilities and securely determine the nature of the accretor.

\begin{figure*}[tb]
  \gridline{
    \fig{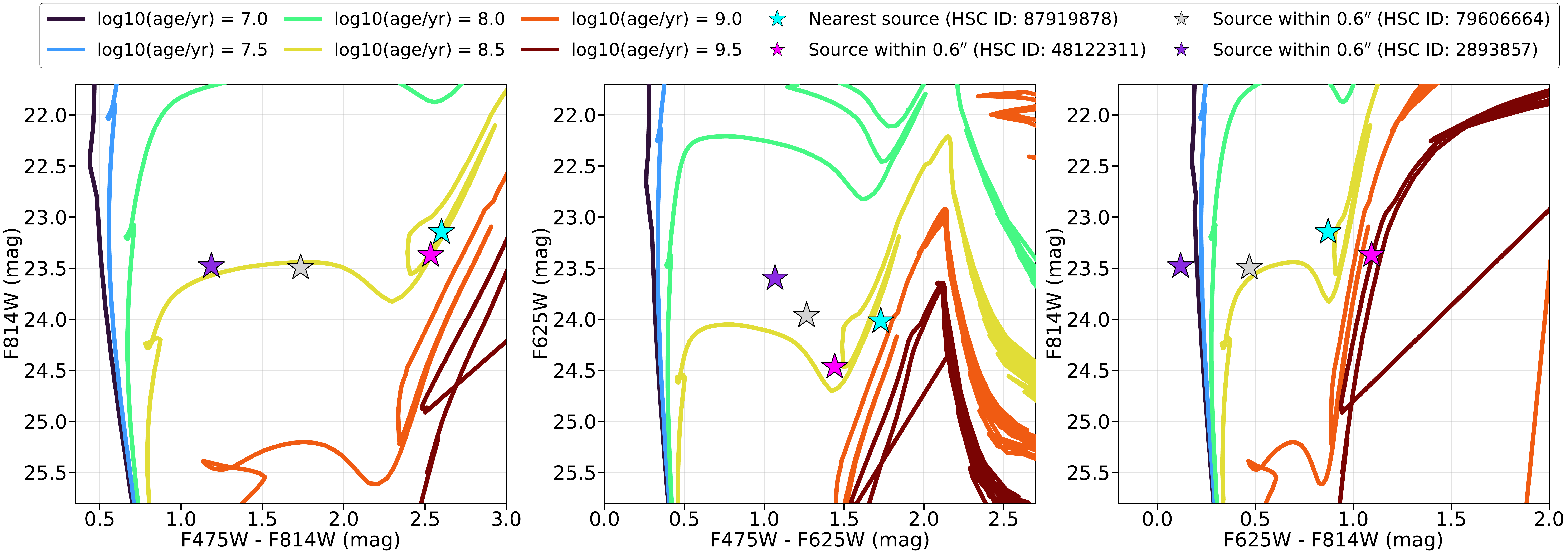}{\textwidth}{}
  }
  \caption{Color--magnitude diagrams constructed from the four HSC sources around M31 ULX-3 that have valid photometric measurements in the F475W, F625W, and F814W bands. The left, middle, and right panels show F814W versus F475W$-$F814W, F625W versus F475W$-$F625W, and F814W versus F625W$-$F814W, respectively. Colored lines represent stellar isochrones with total extinction $A_V=1.58$ for the ages indicated in the legend. The cyan star marks the HSC source nearest to the X-ray position, while the magenta, light gray, and violet stars mark the other sources within $0.6''$.}
  \label{fig:cmd}
\end{figure*}

\subsection{HST Counterparts} \label{subsec:HST_counterparts}
% CMD
% Chance coincidence of the nearest star

\subsubsection{Color--Magnitude Diagrams}

As \textit{HST} covered the region around M31 ULX-3, we searched the HSC for nearby stellar counterpart candidates and their photometry in three filters: F475W, F625W, and F814W. We identified four HSC sources within $0.6''$ of the corrected X-ray position that have valid measurements in all three filters. To characterize the stellar population near M31 ULX-3, we compared the photometry of these sources with isochrones from the PAdova and TRieste Stellar Evolution Code \citep[PARSEC;][]{PARSEC}. We applied the extinction law of \citet{F99_law} to the PARSEC isochrones under the estimated total extinction toward M31 ULX-3, $A_V=1.58$ \citep[]{Wang2025}. Figure~\ref{fig:cmd} presents color--magnitude diagrams (CMDs) for three filter combinations: F814W versus F475W$-$F814W, F625W versus F475W$-$F625W, F814W versus F625W$-$F814W. The nearest HSC source is marked with a cyan star, while the other three HSC sources within $0.6''$ are denoted with a magenta, light gray, and violet stars. The nearest HSC source and the source marked in magenta are broadly consistent with the stellar evolutionary tracks in all three CMDs, whereas the sources marked in light gray and violet show some offset in the F625W$-$F814W color. This may suggest an additional contribution to their observed colors beyond that expected from the stellar isochrones. Such deviations can arise from a range of effects, including emission associated with the accretion disk. An example of similar color offsets is seen in the optical counterpart of the ULX Holmberg~IX~X-1, for which a significant accretion-disk contribution to the optical emission has been suggested \citep{Grise2011}. However, the available photometry does not allow us to determine the origin of the color deviations.

We fitted all four sources by minimizing the residuals across all isochrones in all three filters. We found that the nearest HSC source (HSC ID: 87919878) is best described by an isochrone with $\log_{10}(\mathrm{age/yr})\sim8.5$, corresponding to a stellar mass of $M\sim 3.2\,M_{\odot}$. Intriguingly, the other three HSC sources appear to have the same best-fit age and mass. These inferred properties suggest that the source could be an evolved giant star companion to M31 ULX-3, indicating M31 ULX-3 may be in an intermediate-mass XRB system (IMXB). Alternatively, the M31 ULX-3 may have an undetected lower-mass donor instead, as these interpretations require further confirmation.

\subsubsection{Chance Coincidence Probability}

For each HSC source, we estimated the probability of finding at least one unrelated field source within its angular separation from the X-ray position using Poisson statistics, $P_{\rm chance}=1-\exp(-\pi r^2\Sigma),$ where $r$ is the angular separation between the HSC source and the corrected X-ray position, and $\Sigma$ is the local surface density of HSC sources satisfying a certain photometric criterion. In this work, we used the local surface density extending up to $50''$ from the X-ray position. This metric has been widely used to estimate the probability of chance alignments in counterpart searches and host-galaxy associations for transients and X-ray sources \citep[e.g.][]{Bloom2002_ChanceCoincidence, Ratti2010_ChanceCoincidence, Antoniou2016_ChanceCoincidence, Eftekhari2017_ChanceCoincidence}. 

For a criterion that a field source must be the same or brighter magnitude in all three filters, we found the nearest source (HSC ID: 87919878) to have $P_{\rm chance} \sim20\%$, while the remaining three have $P_{\rm chance} \sim 71\%-85\%$. The higher probabilities for these sources reflect both their larger offsets from the X-ray position and their fainter magnitudes compared to the local field stars around the area. As a more restrictive test, we required the field source to have at least as red or more red colors in each of the three color combinations in Figure~\ref{fig:cmd}: F475W$-$F814W, F475W$-$F625W, and F625W$-$F814W. This gives $P_{\rm chance}\sim1\%$ for nearest source, compared with $P_{\rm chance}\sim10\%-95\%$ for the remaining three. Thus, the nearest source is the least likely to be a chance alignment under either criterion. Nevertheless, these probabilities alone do not establish physical association, and its identification as the companion of M31 ULX-3 remains tentative.

% Although the nearest HSC source is therefore the least likely of the four, its $P_{\rm chance}$ value remains non-negligible. Therefore, its possible identification as a companion to M31 ULX-3 remains speculative.

\subsection{Implications for M31 and Wider ULX Population}

\begin{figure*}[htb!]
  \gridline{
    \fig{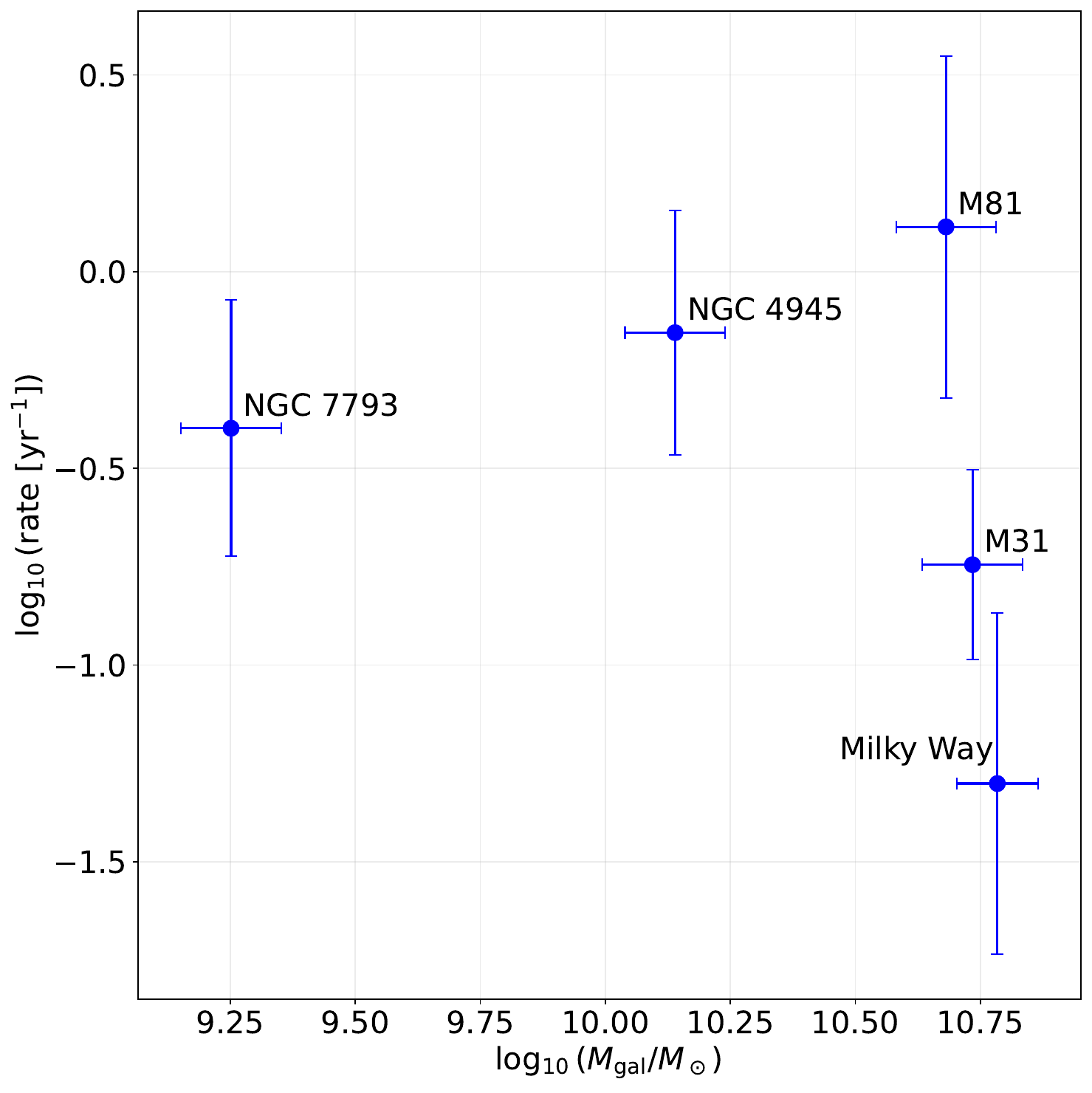}{0.5\textwidth}{}
    \fig{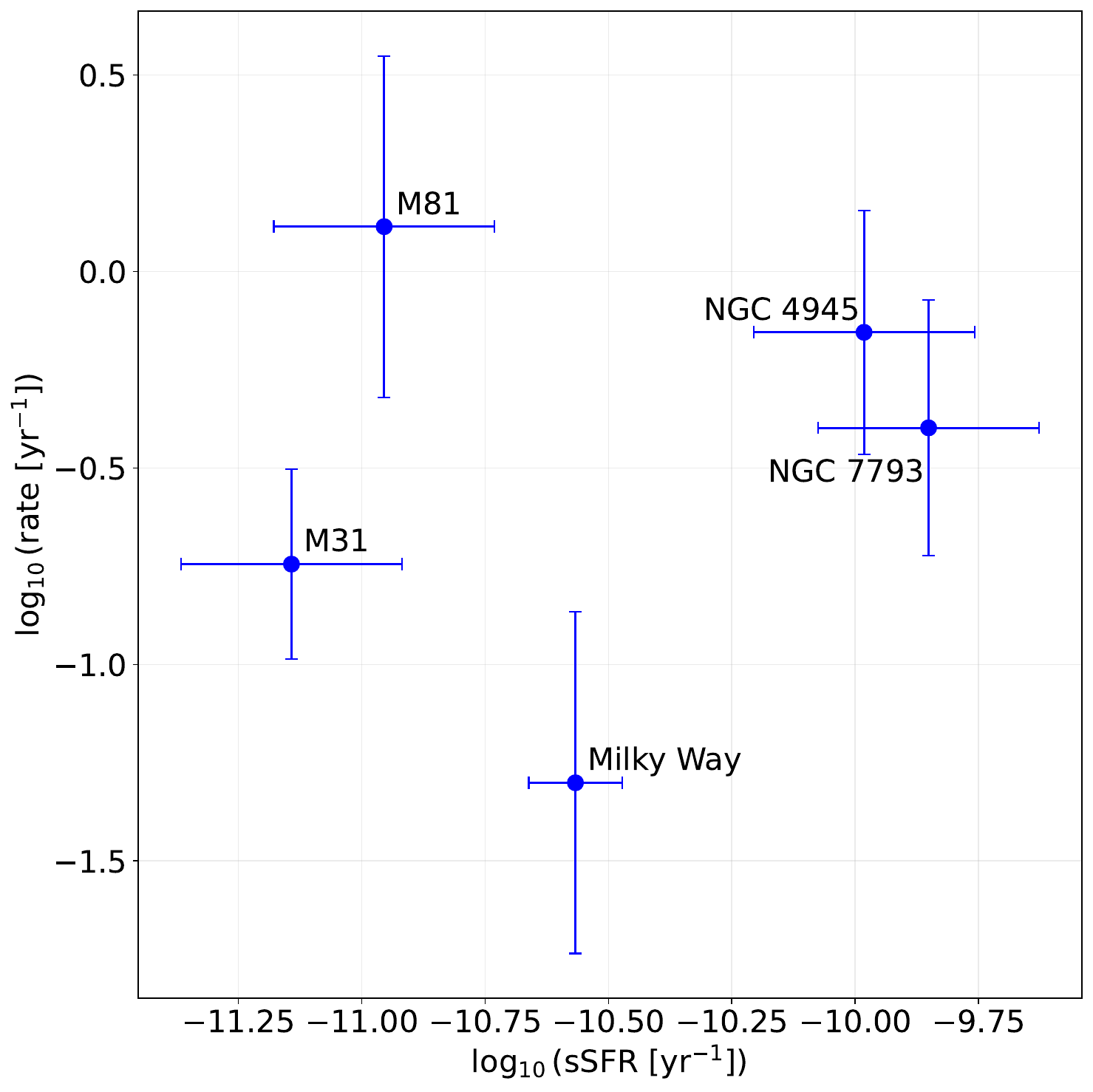}{0.5\textwidth}{}
  }
  \caption{Plot of transient ULX occurrence rate as a function of host-galaxy properties. \textit{Left:} $\log_{10}$ transient ULX occurrence rate versus $\log_{10}(M_{\rm gal}/M_\odot)$. \textit{Right:} $\log_{10}$ transient ULX occurrence rate versus $\log_{10}(\mathrm{sSFR}/{\rm yr}^{-1})$. Points show the Milky Way, M31, M81, NGC 4945, and NGC 7793, with horizontal and vertical uncertainties representing the uncertainties in galaxy properties and transient ULX rates, respectively. The M31 and Milky Way occurrence rates are estimated in this work, while the rates for M81, NGC~4945, and NGC~7793 are taken from \citet{Brightman2023}. Stellar masses and SFRs for M31, M81, NGC~4945, and NGC~7793 are adopted from the z0MGS \citep{Leory2019_z0MGS,Sandstrom2019_z0MGS}, while the corresponding Milky Way properties are taken from \citet{Licquia2015_MW_sSFR_Mass}. The sSFR values are calculated as $\mathrm{SFR}/M_{\rm gal}$.}
  \label{fig:rates}
\end{figure*}

We summarize the spectral properties of M31 ULX-3 obtained with the \texttt{powerlaw} and \texttt{diskpbb} models in Table~\ref{tab:fit_params}. The photon index spans $\Gamma\simeq 1.6 - 2.5$, with a mean value of $\sim2.1$ and a standard deviation of approximately $0.3$. The joint fitted intrinsic absorption, $\rm N_{H,int}$, is approximately $1.05 \times 10^{21} \ \rm cm^{-2}$. This value should not necessarily be interpreted as physical evidence for substantial intrinsic absorption because $\rm N_{H, int}$ is usually inflated in soft powerlaw models $(\Gamma \gtrsim 2)$. Additionally, the \texttt{diskpbb} fits found no significant intrinsic absorption (Section~\ref{subsec:spectral_modeling}). Nonetheless, we found both the fitted photon indices and absorbing columns overlap with those reported for established ULX samples. In particular, \citet{Gladstone2009} found a mean absorbing column of $\rm N_{H}=2.8\times10^{21}\ {\rm cm^{-2}}$, with a standard deviation of $1.7\times10^{21}\ {\rm cm^{-2}}$, and a mean photon index of $\Gamma=2.3$ with a standard deviation of $0.5$. Although the spectral properties of M31 ULX-3 are broadly consistent with other ULXs, the spectral evolution is rarely seen among persistent/transient ULXs.

%The spectral properties of M31 ULX-3 are therefore broadly consistent with the distribution occupied by persistent ULXs, consistent with \citet{Brightman2023} that no significant spectral distinction is currently apparent between transient ULXs and their persistent counterparts.

Based on the three transient ULXs detected in M31 over the constant monitoring of M31 by \textit{Swift}/XRT from 2007 June to 2024 January, we estimate an occurrence rate of $\sim0.18\pm0.10~{\rm yr^{-1}}$. This is lower than the transient-ULX rates reported by \citet{Brightman2023} for NGC 4945 $(0.7\pm0.5 \ \rm yr^{-1})$, NGC 7793 $(0.4\pm0.3 \ \rm yr^{-1})$, and M81 $(1.3 \pm 1.3 \ \rm yr^{-1})$. For comparison, we estimate the Galactic transient ULX occurrence rate, based on Swift~J0243.6+6124, the only confirmed ULX in our Galaxy, discovered by the \textit{Swift Burst Alert Telescope} (\textit{Swift}/BAT) in 2017, to be approximately $0.05\pm0.05 \ \rm yr^{-1}$ over the all-sky X-ray monitoring interval starting from the first \textit{Swift}/BAT observation in 2004 December to 2026 June. 

To explore whether these differences may be related to host-galaxy properties, we compared the transient ULX occurrence rate with the host-galaxy specific star-formation rate (sSFR). Stellar masses $(M_{\rm gal})$ and star-formation rates (SFR) for the external galaxies were adopted from the $z=0$ Multiwavelength Galaxy Synthesis catalog \citep[z0MGS;][]{Leory2019_z0MGS,Sandstrom2019_z0MGS}, while the corresponding Milky Way properties were taken from \citet{Licquia2015_MW_sSFR_Mass}. The sSFR was then calculated as $\mathrm{SFR}/M_{\rm gal}$. 

We examined the dependence of the transient ULX rate on host-galaxy properties by comparing $\log_{10} (\rm rate \ [{\rm yr}^{-1}])$ with both $\log_{10}(M_{\rm gal}/M_\odot)$ and $\log_{10}(\mathrm{sSFR} \ [{\rm yr}^{-1}])$ (Figure~\ref{fig:rates}). A Spearman rank test for the five galaxies mentioned previously gives $\rho=-0.6$ $(p_{\rm spearman}=0.35)$ for the relation with galaxy mass and $\rho=0.1$ $(p_{\rm spearman}=0.95)$ for the relation with sSFR, indicating no statistically significant correlation with either galaxy mass or sSFR. The lack of significant correlations with these global host properties may suggest that other factors, including binary population differences, stellar age distributions, or transient ULX duty cycles, may play an important role in setting the observed occurrence rates. However, these trends should be interpreted cautiously given the small sample size, large measurement errors, and the sensitivity of rank correlations to individual systems. Furthermore, differences in monitoring cadence, sensitivity, and temporal coverage introduce additional uncertainties in the inferred rates, as demonstrated by \citet{Brightman2023} for transient ULX searches using \textit{Swift}.

The detection of M31 ULX-3 adds a third example to the known population of transient ULXs in M31, following M31 ULX-1 and M31 ULX-2. Together with the occurrence rate inferred above, this suggests that transient ULXs may not be exceptionally rare in M31. While M31 ULX-1 and M31 ULX-2 were both interpreted as a probable stellar-mass black hole in  LMXBs \citep{Middleton2012_ULX1, Barnard2013_M31ULX2, Esposito2013,Middleton2013}, M31 ULX-3 may broaden this picture, as our results tentatively favor an IMXB system. Additionally, if future evidence supports M31 ULX-3 as a neutron star, the three systems may indicate that the transient ULX population in M31 is heterogeneous in both compact-object type and donor properties, although this interpretation remains uncertain. 

More broadly, the fact that such a luminous X-ray transient occurred in the nearest neighboring galaxy and remained unnoticed for several years, despite two NASA X-ray telescopes observing M31 at the time, highlights an important limitation of current X-ray transient discovery capabilities. In contrast, a comparably luminous optical transient in M31 would likely have been identified within hours through automated searches and rapid community alerts. M31 ULX-3 therefore illustrates the important value of real-time transient searches and alert infrastructure for X-ray observatories, including those from Chandra and future X-ray missions. In addition, more uniform and sensitive long-term monitoring, coupled with rapid transient-identification pipelines, will be required to both determine whether the transient ULX occurrence rate depends on host-galaxy properties and establish how representative M31 ULX-3 is of the wider transient ULX population.

\section{Conclusion} \label{sec:conclusion}

We report the discovery and X-ray observation and analysis of a newly identified transient ULX in M31: M31 ULX-3 (LSXPS J004321.0+411833). Our main findings are summarized as follows:

\begin{itemize}
    \item M31 ULX-3 is most likely a transient ULX associated with M31. The absence of a bright optical or ultraviolet counterpart, together with the lack of a Gaia DR3 stellar counterpart, disfavors a foreground Galactic origin, including classical novae, XRBs, and magnetar flares. A background origin, such as a tidal disruption event, is also unlikely.

    \item M31 ULX-3 reached a peak $0.3-10$ keV luminosity of $L_{\rm X,peak}\sim 10^{39}\ {\rm erg\ s^{-1}}$ and remained above half of its peak luminosity for $\sim20$ days.

    \item If taken at face value, the spectral evolution of the \texttt{diskpbb} fits suggests a transition from a standard thin disk state at early times toward a broader advective disk at later epochs, opposite to the typical evolution expected during outburst decay. Moreover, the \texttt{powerlaw} fits shows evidence of a transition from a hard to a softer state spectra. This change may indicate the presence of an optically thick wind in the accretion at late times. Nevertheless, one should bear in mind that a good spectral fit does not necessarily imply that the underlying model is correct.
    
    \item The X-ray light curve can be reproduced by the disk-instability model of \citet{Hameury2020}, with characteristic timescales of $\sim41-61$ days. The inferred maximum mass-accretion rate exceeds the Eddington accretion rate for an assumed $1.4\,M_{\odot}$ neutron star model and approximately Eddington for an assumed $10\,M_{\odot}$ stellar-mass black hole.
    
    \item The nearby HSC sources and comparison with stellar isochrones indicate a possible giant donor star with $\log_{10}({\rm age/yr})\sim8.5$, assuming $A_V=1.58$, and an estimated mass of $M\sim3.2\,M_{\odot}$.
    
    \item The inferred transient ULX occurrence rate in M31 is $\sim0.18\pm0.10\ {\rm yr^{-1}}$, which is lower than the rates measured for the three-galaxy sample from \citet{Brightman2023}. A Spearman rank correlation analysis reveals no statistically significant correlation between the transient ULX occurrence rate and host-galaxy properties, although the small number of galaxies and large measurement errors limit the strength of this conclusion.
\end{itemize}

M31 ULX-3 provides an important window to the small but growing population of transient ULXs in M31. Continued, uniform monitoring of M31 and other nearby galaxies will be essential to assess its broader significance.

%% For this sample we use BibTeX plus aasjournalv7.bst to generate the
%% the bibliography. The sample7.bib file was populated from ADS. To
%% get the citations to show in the compiled file do the following:
%%
%% pdflatex sample7.tex
%% bibtext sample7
%% pdflatex sample7.tex
%% pdflatex sample7.tex

\begin{acknowledgements}

    F. acknowledges support from the Caltech Summer Undergraduate Research Fellowship (SURF) program. D.J.W. acknowledges support from the Science and Technology Facilities Council (STFC; grant code ST/Y001060/1). M.J.M. acknowledges support by the Science and Technology Facilities Council (STFC) via small grant award ST/Y001699/1. W.J.-G. is supported by NASA through Hubble Fellowship grant HSTHF2-51558.001-A awarded by the Space Telescope Science Institute, which is operated for NASA by the Association of Universities for Research in Astronomy, Inc., under contract NAS5-26555.
    
    This research has made use of data and/or software provided by the High Energy Astrophysics Science Archive Research Center (HEASARC), which is a service of the Astrophysics Science Division at NASA/GSFC. This research has made use of the NuSTAR Data Analysis Software (NuSTARDAS) jointly developed by the ASI Space Science Data Center (SSDC, Italy) and the California Institute of Technology (Caltech, USA). This research has made use of the XRT Data Analysis Software (XRTDAS) developed under the responsibility of the ASI Space Science Data Center (SSDC), Italy. This work made use of data supplied by the UK Swift Science Data Centre at the University of Leicester. 

    This paper employs a list of Chandra datasets, obtained by the Chandra X-ray Observatory, contained in the Chandra Data Collection~\dataset[DOI: 10.25574/cdc.647]{https://doi.org/10.25574/cdc.647}.

    Some of the data presented in this paper were obtained from the Mikulski Archive for Space Telescopes (MAST) at the Space Telescope Science Institute. Based on observations made with the NASA/ESA Hubble Space Telescope, and obtained from the Hubble Legacy Archive, which is a collaboration between the Space Telescope Science Institute (STScI/NASA), the Space Telescope European Coordinating Facility (ST-ECF/ESAC/ESA) and the Canadian Astronomy Data Centre (CADC/NRC/CSA). The Hubble Source Catalog can be accessed via~\dataset[DOI: 10.17909/T97P46]{https://doi.org/10.17909/T97P46} and the specific observations are associated with programs: GO12109 and GO13710, and can be accessed via~\dataset[DOI: 10.17909/1yj3-j262]{https://doi.org/10.17909/1yj3-j262}. STScI is operated by the Association of Universities for Research in Astronomy, Inc., under NASA contract NAS5–26555. Support to MAST for these data is provided by the NASA Office of Space Science via grant NAG5–7584 and by other grants and contracts.

    The Digitized Sky Survey was produced at the Space Telescope Science Institute under U.S. Government grant NAG W–2166. The images of these surveys are based on photographic data obtained using the Oschin Schmidt Telescope on Palomar Mountain and the UK Schmidt Telescope. The plates were processed into the present compressed digital form with the permission of these institutions.

    This work has made use of data from the Asteroid Terrestrial-impact Last Alert System (ATLAS) project. The Asteroid Terrestrial-impact Last Alert System (ATLAS) project is primarily funded to search for near-earth asteroids through NASA grants NN12AR55G, 80NSSC18K0284, and 80NSSC18K1575; byproducts of the NEO search include images and catalogs from the survey area. This work was partially funded by Kepler/K2 grant J1944/80NSSC19K0112 and HST GO-15889, and STFC grants ST/T000198/1 and ST/S006109/1. The ATLAS science products have been made possible through the contributions of the University of Hawaii Institute for Astronomy, the Queen’s University Belfast, the Space Telescope Science Institute, the South African Astronomical Observatory, and The Millennium Institute of Astrophysics (MAS), Chile.

    We acknowledge the use of ChatGPT (GPT-5.6 Terra) for language editing to improve the wording and readability of portions of the manuscript. ChatGPT was also used to assist in refining certain figures presented in this work.
\end{acknowledgements}

\facilities{CXO, Swift (XRT, UVOT), XMM-Newton, NuSTAR, ATLAS, Gaia, PO:1.2m, HST}
\software{\texttt{Astropy} \citep{Astropy2013, Astropy2018, Astropy2022}, \texttt{CIAO} \citep{CIAO2006}, \texttt{emcee} \citep{emcee}, \texttt{HEASoft} \citep{HEASoft2014}, \texttt{NumPy} \citep{NumPy2020}, \texttt{Pandas} \citep{Pandas2010, Pandas2020}, \texttt{PyXspec} \citep{PyXspec}, \texttt{SAOImage DS9} \citep{DS9_2003}, \texttt{SciPy} \citep{SciPy2020}, \texttt{XSPEC} \citep{xspec1996}, \texttt{XSNAP} \citep{Ferdinand2026}}

\bibliography{main}{}
\bibliographystyle{aasjournalv7}

\appendix
\vspace{-0.25cm}
\restartappendixnumbering

\section{Best-fitting Parameters of M31 ULX-3}

Table~\ref{tab:fit_params} reports the best-fitting parameters of the \texttt{powerlaw} model and \texttt{diskpbb} model. All observations are taken from \textit{Swift}/XRT and fitted with a tied intrinsic absorption across all epoch groups. The resulting tied $\rm N_{H, int}$ is $= 1.05^{+0.30}_{-0.28} \times 10^{21} \ \rm cm^{-2}$ for the \texttt{powerlaw} model and $1.22 \times 10^{16} \ \rm cm^{-2}$ with a $3\sigma$ upper limits of $\sim7.7 \times 10^{20} \ \rm cm^{-2}$ for the \texttt{diskpbb} model.

\begin{deluxetable}{cccccccc}[!h]
\setlength{\tabcolsep}{10pt}
\tabletypesize{\small}
\tablewidth{0pt}

\tablecaption{Best-fitting Parameters of M31 ULX-3
\label{tab:fit_params}}

\tablehead{
\colhead{} &
\colhead{} &
\multicolumn{2}{c}{\texttt{powerlaw} Model} &
\colhead{} &
\multicolumn{3}{c}{\texttt{diskpbb} Model}
\\
\cline{3-4}
\cline{6-8}
\colhead{Mid Time\tablenotemark{a}} &
\colhead{Observation ID} &
\colhead{$\Gamma$} &
\colhead{$\log_{10}(\rm Flux)$\tablenotemark{b}} &
\colhead{} &
\colhead{$p$} &
\colhead{$T_{\rm in}$} &
\colhead{$\log_{10}(\rm Flux)$\tablenotemark{b}} 
\\[-5pt]
\colhead{(days)} &
\colhead{} &
\colhead{} &
\colhead{(Unabsorbed)} &
\colhead{} &
\colhead{} &
\colhead{(keV)} &
\colhead{(Unabsorbed)} 
}

\startdata
3.22 & 00013977006, 00013977007 & $1.56^{+0.10}_{-0.10}$ & $-10.70^{+0.03}_{-0.03}$ & & $0.76^{+0.11}_{-0.07}$ & $1.48^{+0.29}_{-0.20}$ & $-10.85^{+0.04}_{-0.03}$ \\[3pt]
8.16 & 00013977008, 00032159019 & $1.80^{+0.09}_{-0.09}$ & $-10.87^{+0.02}_{-0.02}$ & & $0.60^{+0.03}_{-0.02}$ & $2.04^{+0.53}_{-0.33}$ & $-10.96^{+0.03}_{-0.03}$ \\[3pt]
19.25 & 00013977009, 00032159020 & $2.07^{+0.11}_{-0.10}$ & $-11.16^{+0.03}_{-0.02}$ & & $0.57^{+0.03}_{-0.03}$ & $1.52^{+0.41}_{-0.21}$ & $-11.28^{+0.03}_{-0.03}$ \\[3pt]
26.18 & 00014412001, 00014412002, & $2.12^{+0.15}_{-0.15}$ & $-11.25^{+0.04}_{-0.04}$ & & $0.58^{+0.06}_{-0.05}$ & $1.27^{+0.48}_{-0.23}$ & $-11.41^{+0.05}_{-0.04}$ \\[3pt]
27.97 & 00014412003, 00014412005, & $2.11^{+0.15}_{-0.15}$ & $-11.34^{+0.04}_{-0.04}$ & & $0.58^{+0.06}_{-0.05}$ & $1.39^{+0.60}_{-0.38}$ & $-11.48^{+0.05}_{-0.04}$ \\[-3pt]
& 00013977010 & & & & & & \\[3pt]
35.90 & 00014412006, 00014412007, & $2.04^{+0.13}_{-0.13}$ & $-11.40^{+0.03}_{-0.03}$ & & $0.64^{+0.09}_{-0.06}$ & $1.05^{+0.22}_{-0.15}$ & $-11.59^{+0.03}_{-0.04}$ \\[-3pt]
& 00014412008, 00014412012, & & & & & & \\[-3pt]
& 00014412013, 00014412014, & & & & & & \\[-3pt]
& 00032159022, 00014412015, & & & & & & \\[-3pt]
& 00013977011 & & & & & & \\[3pt]
119.50 & 00014412019, 00014412020, & $2.73^{+0.22}_{-0.21}$ & $-12.06^{+0.06}_{-0.06}$ & & $0.52^{+0.09}_{-0.02}$ & $0.79^{+28}_{-0.18}$ & $-12.29^{+0.06}_{-0.06}$ \\[-3pt]
& 00013977012, 00013977013, & & & & & & \\[-3pt]
& 00013977014, 00013977015, & & & & & & \\[-3pt]
& 00013977017, 00013977019, & & & & & & \\[-3pt]
& 00013977020, 00013977021 & & & & & & 
\\[3pt]
203.38 & 00013977022, 00013977024, & $2.18^{+0.26}_{-0.25}$ & $-11.84^{+0.06}_{-0.06}$ & & $0.57^{+0.11}_{-0.07}$ & $1.28^{+6.54}_{-0.39}$ & $-11.99^{+0.12}_{-0.07}$ \\[-3pt]
& 00013977025 & & & & & & \\
\enddata
\tablecomments{All errors are reported at $1\sigma$ confidence level (corresponding to $\Delta C=1.0$)}
\tablenotetext{a}{With respect to MJD 59374}
\tablenotetext{b}{$\log_{10}$(Flux) is from 0.3–10 keV. Flux units in $\rm erg \ cm^{-2} \ s^{-1}$}
\end{deluxetable}

\section{Swift/XRT \texttt{Diskpbb} Spectra Fits Plots}

\begin{figure*}[htb!]
  \gridline{
    \fig{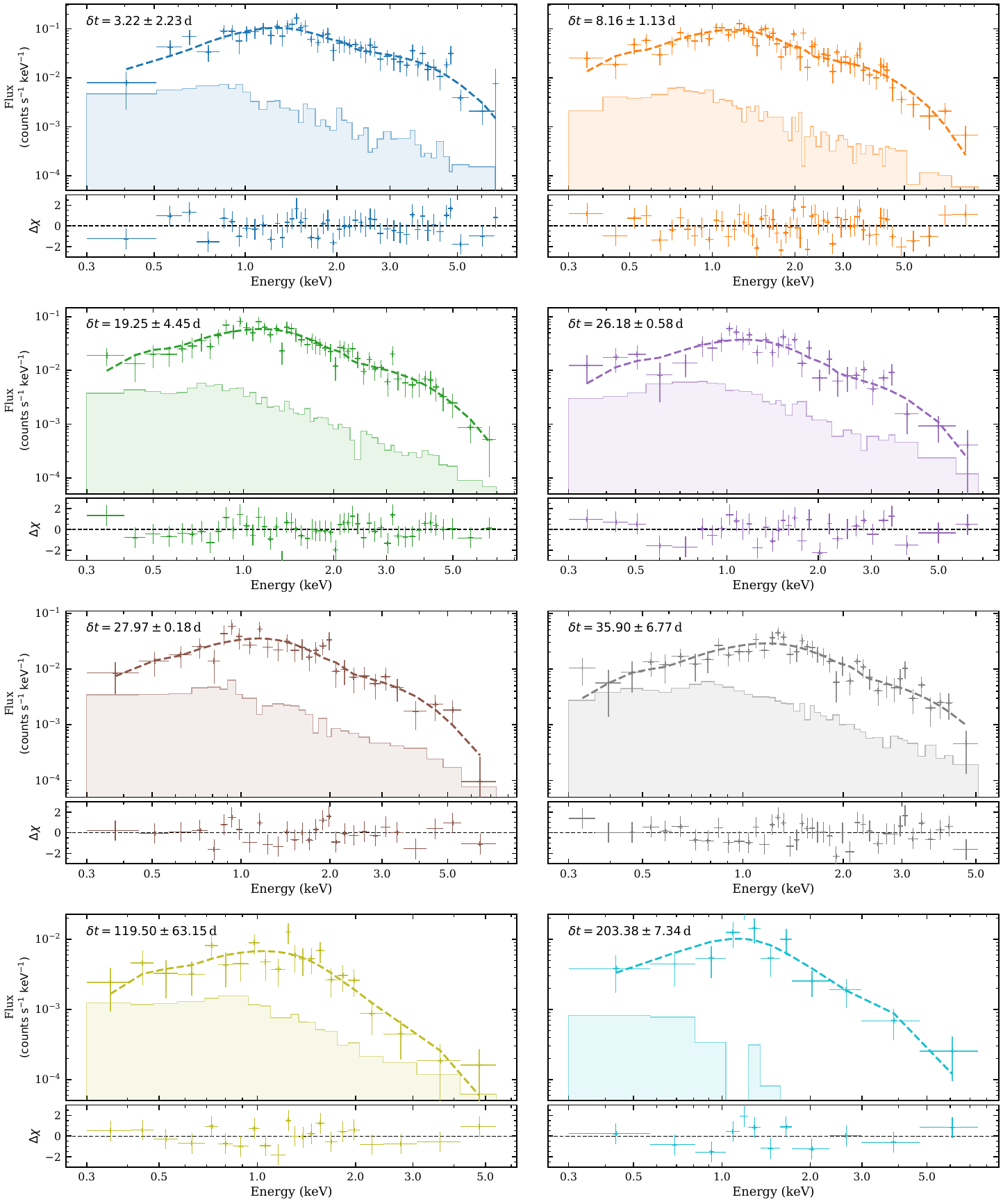}{0.95\textwidth}{}
  }
  \caption{Plot of M31 ULX-3 spectra data (above) and $\Delta \chi$ (bottom) for all 8 \textit{Swift}/XRT epoch groups. Histogram represents the background data and a best-fit \texttt{diskpbb} model is shown in a dashed line for each corresponding epoch group.}
  \label{fig:spectra_data}
\end{figure*}

Figure~\ref{fig:spectra_data} shows the spectra data fits of the \texttt{diskpbb} model.

%% This command is needed to show the entire author+affiliation list when
%% the collaboration and author truncation commands are used.  It has to
%% go at the end of the manuscript.
%\allauthors

%% Include this line if you are using the \added, \replaced, \deleted
%% commands to see a summary list of all changes at the end of the article.
% \listofchanges

\end{document}